\documentclass[fleqn,usenatbib]{mnras}

\usepackage{graphicx}	
\usepackage{amsmath}	
\usepackage{amssymb}	
\usepackage{multicol}        
\usepackage{bm}		
\usepackage{pdflscape}	
\usepackage{hyperref}
\usepackage{amsfonts}
\usepackage{txfonts}
\usepackage{ctable}
\usepackage{url}
\usepackage[normalem]{ulem}

\usepackage{color}
\usepackage{grffile}

\newcommand{\kms}{\,km\,s$^{-1}$} 

\newcommand\altaffilmark[1]{$^{#1}$}
\newcommand\altaffiltext[1]{$^{#1}$}
\newcommand{\driftvelmag}{w_{\rm s}}

\newcommand{\larmorfreq}{\omega_{\rm L}}

\newcommand \bB         {{\bf B}}

\newcommand \bE         {{\bf E}}

\newcommand \beq        {\begin{equation}}
\newcommand \beqa	{\begin{eqnarray}}

\newcommand \eeq	{\end{equation}}
\newcommand \eeqa	{\end{eqnarray}}

\newcommand \HH	        {{\rm H}_2}

\newcommand \K  	{\,{\rm K}}

\newcommand \muG        {\mu{\rm G}}

\newcommand \um         {$\mu${\rm m}}

\newcommand{\dvel}{{\bm v}}
\newcommand{\gvel}{{\bm u}}

\newcommand{\drift}{{\bm w}_{\rm s}}
\newcommand{\agr}{a_{\rm gr}}

\newcommand \pr {{Phys.~Rev.}}

\newcommand{\oldtext}[1]{}

\usepackage[T1]{fontenc}
\usepackage{ae,aecompl}

\title[Grain dynamics I]{Dust dynamics in RAMSES - I. Methods and turbulent acceleration
 \vspace{-0.5cm}}

\author[Moseley et al.]{
\parbox[t]{\textwidth}{ 
	Eric R.~Moseley\altaffilmark{1}\thanks{E-mail: moseley@princeton.edu},
	R. Teyssier\altaffilmark{1},
	B.~T. Draine\altaffilmark{1}
} 
\vspace*{6pt} \\
\altaffiltext{1}{Department of Astrophysical Sciences, Princeton University, Princeton, NJ 08540, USA}\\
}

\date{}

\pubyear{2022}

\begin{document}
\maketitle

\begin{abstract} 
Supernova ejecta and stellar winds are believed to produce interstellar dust grains with relatively large sizes. Smaller grains can be produced via the shattering of large grains that have been stochastically accelerated. To understand this stochastic acceleration, we have implemented novel magnetohydrodynamic(MHD)-particle-in-cell(PIC) methods into the astrophysical fluid code RAMSES. We treat dust grains as a set of massive ``superparticles'' that experience aerodynamic drag and Lorentz force. We subject our code to a range of numerical tests designed to validate our method in different physical conditions, as well as to illustrate possible mechanisms by which grains can be accelerated. As a final test as well as a foundation for future work, we present the results of decaying dusty MHD turbulence simulations with grain parameters chosen to resemble 1-2 $\mu$m grains in typical cold neutral medium conditions. We find that in these conditions, these grains can be effectively accelerated to well beyond their shattering velocities. This is true for both electrically charged and neutral grains. While the peak of the gas-grain relative drift velocity distribution is higher for neutral grains, the drift velocity distribution of charged grains exhibits an extended exponential tail out to much greater velocities. Even so, the shapes of the distributions are such that the extra gas-grain coupling provided by the Lorentz force offers grains relative protection from shattering. We also discuss the connection between our simulations and the relatively pristine $\sim\mu{\rm m }$ sized presolar grains that do not appear to have undergone significant wear in their lifetimes.
\end{abstract}

\section{Introduction}\label{sec:intro}
\subsection{Overview}
First discovered by \citet{Trumpler_1930}, astrophysical dust plays a fundamental role in planet, star, and galaxy formation and evolution {\citep{hirashita2002effects, hofner2018mass}, gas cooling and chemical evolution \citep{Draine_2003a}, as well as in stellar and active galactic nuclear ``feedback'' processes \citep{Thompson+Quataert+Murray_2005}. These feedback processes are dependent in large part on the interaction between dust and radiation.} Dust reprocesses perhaps 30\% of all stellar radiation into the infrared \citep{bernstein2002first}. {For individual galaxies \citep[e.g. LIRGs,][]{soifer1984infrared}, this fraction can be much higher. Within galaxies, this reprocessing means that where radiation is anisotropic, dust-driven winds are possible with dramatic implications for the transport of galactic energy and momentum, as well as the rate of star formation \citep{sandford1984radiatively, chang1987effect, franco1991photolevitation, berruyer1991dust, Lamers_Cassinelli_1999}. }

{Even ignoring the impact of dust in launching winds, dust is both chemically and energetically important at multiple length scales in the interstellar medium (ISM). In addition to being a reservoir of metals, dust grains are usually the main site of formation for molecular hydrogen ($\HH$), even in metal-poor galaxies or galactic regions \citep{Hirashita+Hunt+Ferrara_2002,Draine_2011a}. $\HH$ has rovibrational states that act as strong cooling channels for  warm gas, even being the dominant cooling source at some densities and temperatures \citep[e.g.][]{Moseley+Draine+Tomida_2021}. As $\HH$ forms on the surfaces of dust grains primarily, the amount of available surface area on grain surfaces is an important quantity. This depends partly on the size distribution of dust grains as inferred by line-of-sight extinction curves \citep{Mathis+Rumpl+Nordsieck_1977,Weingartner+Draine_2001a}. Thus, both local enhancement of the dust-to-gas mass ratio as well as variations in the grain size distribution are relevant to the local $\HH$ production rate.}

Both large and small grains are of dynamical interest: large grains ($a_{\rm gr} \gtrsim 0.1\,\mu{\rm m}$) possess most of the dust mass, while small grains ($a_{\rm gr} \lesssim 0.1\,\mu{\rm m}$) possess most of the cross-sectional area \citep{Mathis+Rumpl+Nordsieck_1977}. 

When concentrated dynamically, large grains may posses non-negligible mass both inertially and gravitationally with implications for star and planet formation \citep[e.g.][]{bai2010particle, Chiang+Youdin_2010,zhuravlev2021dynamic}.
{The processes that can concentrate large grains are myriad and still being understood. Among them are the ``streaming instability'' \citep{youdin2005streaming}, turbulent concentration of grains \citep{hopkins2016fundamentally,lee2017dynamics,tricco2017dust,Beitia+Gomez_2021,Beitia+Gomez+Vallejo_2021}, and dust ``traps'' \citep[e.g. pressure bumps, vorticies][]{Johansen_2014,Mattson+Hedvall_2022,Haugen+Brandenburg+Mattson_2022}. For more than a decade now, it has been thought that these processes might help to overcome the problem of how centimeter-sized ``pebbles'' can agglomerate into kilometer-sized ``boulders'', thus making possible the formation of planetesimals, but whether or not this is the case is not yet clear \citep[][]{johansen2009particle,yang2017concentrating,Carrera+Simon_2022}.}

{Small grains, on the other hand, having fractionally more surface area than large grains, serve as important molecular formation sites (e.g. $\HH$). Small grains seem to be fractionally less abundant than large grains at high redshift ($z\gtrsim6$) on both theoretical and observational grounds \citep{Aoyama+Hirashita_2020, Di_2021}. One explanation for this is that grains are created at relatively large sizes in type II supernovae \citep{Todini+Ferrara_2001}. These large grains then collide with one another, shattering to produce smaller grains at lower redshifts. Understanding the details of this shattering process depends upon an understanding of the way dust grains interact with MHD turbulence.}
All of the aforementioned factors motivate the development of codes that model dust dynamically. There have been various approaches to this problem, each with different strengths and weaknesses depending upon the regime of interest. One is the ``terminal velocity'' approximation, whereby dust is assumed to have zero inertia, moving exactly at its equilibrium velocity at all times \citep{laibe2014dusty, tricco2017dust, lebreuilly2019small}. This allows for a single-fluid diffusion treatment of dust and gas, where the dust-to-gas mass ratio is simply an additional variable that is advected according to its own equations of motion. Other authors have assumed a two-fluid model \citep[e.g.][]{laibe2012dusty, laibe2012dustyII}. 

On small scales where the dust stopping length (roughly the distance for transsonic dust to slow down by a factor of $\sim2$) is non-negligible, it is often better to treat dust kinetically while the gas is treated with the standard equations of hydrodynamics or magnetohydrodynamics (MHD) \citep{youdin2007protoplanetary, carballido2008kinematics, bai2010particle, pan2011turbulent, hopkins2016fundamentally, lee2017dynamics, steinwandel2021optical}. 
This is the same method that is known in the cosmic-ray literature as MHD-particle-in-cell (PIC) \citep{bai2015magnetohydrodynamic,bai2019magnetohydrodynamic, mignone2018particle, dominguez2021morphology}. In MHD-PIC codes, the kinetic component (whether dust or cosmic rays) is treated as a collection of ``superparticles'', each of which represents a cloud of particles with the same velocity, mass, charge, and other relevant microscopic properties. This cloud has a finite spatial extent determined by a kernel of a given shape. In grid-based codes, this kernel is usually either a uniform cube (known as cloud-in-cell (CIC)), or a `triangle-shaped-cloud' (TSC). The trajectories of the superparticles are then followed in real-time, and the mass, momentum, and energy of these particles are included in the subsequent calculations self-consistently. 

{While the MHD-PIC approach has utility in the study of all of the aforementioned problems, the problem of how grains are stochastically accelerated and subsequently shattered is one for which it is particularly well-suited. As mentioned previously,} dust grains are thought to be produced at large sizes by supernova explosions and in stellar outflows \citep{gehrz1989sources,gail1999mineral, plane2013nucleation}. These initial grains may grow even larger through accretion of gas-phase metals \citep{Spitzer_1978}, or through grain coagulation \citep{Chokshi+Tielens+Hollenbach_1993}. The large dust grains are imperfectly coupled to the underlying gas, and through various stochastic processes accelerated to high relative velocities \citep{Lazarian+Yan_2002, Yan+Lazarian_2003, Yan+Lazarian+Draine_2004, Hirashita+Yan_2009}. Once accelerated, these large grains are shattered through grain-grain collisions, producing small grains \citep{Jones+Tielens+Hollenbach_1996}. Possessing the majority of the cross-sectional area, small grains are important for the obscuration and reprocessing of starlight, as well as the gas-grain coupling that leads to dust-driven winds.
{Understanding the shattering of large grains can also help us to understand the upper end of the grain-size distribution, which may be of importance to star and planet formation and fluctuations in metallicity. This problem has motivated us to implement dust MHD-PIC methods into the astrophysical fluid code RAMSES \citep{teyssier2002cosmological, teyssier2006kinematic, fromang2006high}.}

\subsection{A guide to this paper}\label{sec:guide}

In Section~\ref{sec:methods} of this paper, we describe the methods we have developed for the study of dust-gas mixtures within the finite-volume based framework provided by RAMSES. The methods we employ combine elements of those developed by other authors, such as the superparticle method \citep[e.g.][]{youdin2007protoplanetary, bai2010particle} and particle-mesh back-reaction \citep{yang2016integration}, which enables integration of the system cell-by-cell. In sections~\ref{sec:lorentz} and \ref{sec:drag}, we describe our operator-split method for integrating drag and Lorentz forces on grains within each cell. We have implemented three separate drag laws into our code: linear drag that is independent of gas density, gas density dependent drag \citep[Epstein,][]{epstein1924resistance}, and drag which depends both on the gas density and non-linearly on the relative gas-grain velocity \citep[Epstein-Baines,][]{Baines+Williams+Asebiomo_1965, kwok1975radiation, Draine+Salpeter_1979a}. Our methods give reasonable results, even when the timestep is shorter than the dust stopping and Larmor times, or when the dust-to-gas mass ratio exceeds unity. These stability properties are possible because we treat the drag fully implicitly while treating the Lorentz forces semi-implicitly. 

In Section~\ref{sec:results}, we present the results of a number of numerical tests, the results of which not only serve to validate our code, but also to build physical understanding of the process of dust acceleration. These tests are general, and can be used to test the stability and performance of codes that model massive dust grains subject to drag and Lorentz forces. All tests are performed assuming dust grains all share the same charge, mass, etc. Section~\ref{sec:gyrodamp} shows our simplest test, bulk gyromotion of a homogeneous dust-gas medium. This test builds on the DUSTYBOX test of \citet{laibe2012dusty} by adding Lorentz forces. In Section~\ref{sec:dusty_shocks}, we show the results of simulations of a shock propagating into a uniform dusty medium. We perform this test at varying grain charge-to-mass ratios, dust-to-gas mass ratios, and grain stopping times. Then, in Section~\ref{sec:dusty_alfven}, we present the results of simulations of a circularly polarized Alfv{\'e}n wave in a medium where dust is initially at rest with respect to the mean gas flow. We vary the dust-to-gas mass ratio and charge-to-mass ratio and observe the impact that this has on dust gyroresonant acceleration. This test is interesting because dust gyroresonance with Alfv{\'e}n waves has been proposed as an dominant mechanism in accelerating dust grains to high velocities in a number of interstellar environments, preferentially promoting the shattering of large grains \citep{Yan+Lazarian+Draine_2004}.

Section~\ref{sec:orszag-tang} shows the results of dusty Orszag-Tang vortex simulations \citep{orszag1979small} and compare these results to the simple shock tests in Section~\ref{sec:dusty_shocks} in order to qualitatively explain the observed features. This test serves as a useful bridge between the simplistic one-dimensional simulations in sections~\ref{sec:dusty_shocks}, \ref{sec:dusty_alfven} and the more complex three-dimensional simulations we present later on.

In order to demonstrate the functionality of our dust momentum feedback (back-reaction) implementation, in Section~\ref{sec:rdi} we reproduce the results of the simulation of the magnetized resonant drag instability (RDI) shown in \citet[][hereafter SHS19]{seligman2019non}. We describe the physics behind this instability in more detail in Section~\ref{sec:rdi}. This test is important because relatively few codes have thus far implemented both Lorentz forces on dust grains and back-reaction simultaneously. Thus, reproducing the results obtained with the multi-method code GIZMO \citep{Hopkins_2015} is a useful check. {In addition, the phenomenology of the magnetized RDI is richer than the ordinary acoustic RDI, as there are far more available modes with which the dust may interact. Furthermore,} as the magnetized RDI only grows for dust-to-gas mass ratio $\mu > 0$,  the growth of this instability is a useful test of our implementation of dust back-reaction.\footnote{{Simulations that neglect back-reaction are effectively $\mu=0$ simulations.}}

Finally, in Section~\ref{sec:decayturb} we show the results of high resolution dust-laden decaying MHD turbulence simulations meant to reflect typical conditions in diffuse molecular clouds. We use these simulations to examine both dust density fluctuations and grain acceleration. To our knowledge, this is the first time that MHD-PIC simulations have been used to study dust acceleration specifically. 
The dust grains we model in these simulations are large, 1-2\um$\,$ in size depending upon the composition, meaning that they have stopping-lengths and gyro-radii that are resolvable at the resolution we use. We run simulations with and without Lorentz forces on dust grains, and with two different drag laws to illustrate the effect that these variables have on the dust grain velocity probability density functions (PDFs), and by extension, on grain shattering. We conclude that charged grains of this size in this environment are significantly better-coupled to the gas than uncharged grains and marginally less likely to undergo shattering. Despite this average behavior, we find that simulations with charged grains show a small population of grains in an exponential tail out to much higher velocities. These results motivate further study into the velocity distributions of large grains in diffuse molecular clouds \citep[][in prep.]{MoseleyInPrep}. We use our simulation results to estimate the mean lifetimes of large micron sized grains in the ISM implied by the present work. 

In Section~\ref{sec:discussion}, we discuss our results in the context of other work, such as the gyro-resonant theory of dust acceleration presented in \citet{Yan+Lazarian+Draine_2004} and the models of dust in shocks by \citet{guillet2009shocks}. We also discuss how our results for variation of the dust-to-gas mass ratio compare to those from \citet{hopkins2016fundamentally, lee2017dynamics}. Finally, we discuss how our estimated grain lifetimes compare to those estimated by studies of micron-sized presolar grains trapped in meteorites \citep{Takigawa_2018, Heck_2020}. We provide a summary of this study in Section~\ref{sec:summary}.




\section{Methods}\label{sec:methods}
We model dust with the ``superparticle'' method \citep{youdin2007protoplanetary, carballido2008kinematics,bai2010particle, pan2011turbulent}, where each Lagrangian dust particle represents an ensemble of grains with the same velocity, size, and charge distributed over a small volume by a weighting kernel.
\subsection{Equations solved} \label{sec:equations}
The velocity of a single dust grain evolves according to
\begin{align}
    \frac{{\rm d}\dvel}{{\rm d}t} &= {\bm a}_{\rm gas-dust} + {\bm a}_{\rm ext}\nonumber,\\
    &=-\nu_{\rm s}(\dvel-\gvel) + \omega_{\rm L}(\dvel-\gvel)\times\hat{\bm b}+ {\bm a}_{\rm ext,dust} , \label{eq:dustvel}
\end{align}
where $\frac{{\rm d}}{{\rm d}t}$ is the Lagrangian derivative, $\dvel$ is the grain velocity, $\gvel$ is the fluid velocity, $\nu_{\rm s}$ is the stopping rate, $\omega_{\rm L}$ is the Larmor frequency, $\hat{\bm b}$ is the unit vector along the direction of the magnetic field, ${\bm a}_{\rm ext,dust}$ is the acceleration that dust grains experience due to external forces (e.g. gravity, radiation), and ${\bm a}_{\rm gas-dust}$ is the acceleration grains experience as a direct result of forces exerted on them by the gas. We have implemented three different options for the stopping rate: a constant stopping rate, Epstein drag, and Epstein drag including the supersonic Baines correction \citep{epstein1924resistance,Baines+Williams+Asebiomo_1965, kwok1975radiation, Draine+Salpeter_1979a}:
\begin{align}
    \nu_{\rm s} &= \nu_{0}, \hspace{0.5cm}&({\rm constant})\label{eq:constant_drag}\\
    \nu_{\rm s} &= \sqrt{\frac{8}{\pi\gamma}}\varepsilon^{-1}\frac{\rho c_{\rm s}}{\rho_0 \ell_0},&({\rm Epstein})\label{eq:epstein}\\
    \nu_{\rm s} &= \sqrt{\frac{8}{\pi\gamma}}\varepsilon^{-1}\frac{\rho c_{\rm s}}{\rho_0 \ell_0}\bigg(1+\frac{9\pi\gamma}{128}\frac{\drift^2}{c_{\rm s}^2}\bigg)^{1/2}.&({\rm Epstein-Baines})
    \label{eq:kwok}\\
    \varepsilon &= \frac{\rho_{\rm d}^{\rm i} a_{\rm gr}}{\rho_0 \ell_0}.\label{eq:grain_size}
\end{align}
We will often refer to $t_{\rm s} \equiv \nu_{\rm s}^{-1}$ as the dust stopping time. 

{While a constant drag coefficient is not physically meaningful, it is useful insofar as it can be used in various numerical tests to more easily understand the results and compare against analytic solutions.}

We will also often refer to the dimensionless charge-to-mass ratio $\xi$ rather than $\omega_{\rm L}$. This value can be written as,
\begin{equation}
    \xi \equiv \omega_{\rm L} t_A = \frac{Z_{\rm d} e}{m_d c}\ell_0\sqrt{4\pi\rho_0} .\label{eq:ctm}
\end{equation}
Here, $Z_{\rm d}$ is the charge on a dust grain in units of the fundamental charge $e$, $m_d$ is the mass of a single dust grain, $c$ is the speed of light, and{ $t_{\rm A}$ is the Alfv{\'e}n crossing time (for mean simulation parameters). We can see that $\xi$ is essentially the number of Larmor periods in an Alfv{\'e}n crossing time up to a factor of $2\pi$.\footnote{$\xi$ is also related to the grain `charge parameter' from SHS19 by $\bar{\phi}_{\rm d} \equiv \varepsilon\xi.$ SHS19's charge parameter $\bar{\phi}_{\rm d}$ is directly proportional to the `magnetization' of grains, $\omega_{\rm L} t_{\rm s}$.} When the grain equations of motion are written out in a dimensionless form (as they are in most hydrodynamical codes), $\xi$ and $\varepsilon$ act together to relate the intrinsic properties of the grain to their large-scale dynamical motions in the simulation.}

The gas evolves according to a modified version of the Euler-magnetohydrodynamic (MHD) equations that includes the ``back-reaction'' from the dust back onto the gas:
\begin{align}
    \rho\frac{{\rm d}\gvel}{{\rm d}t} &= -\nabla\bigg( P + \frac{B^2}{8\pi}\bigg) + \frac{(\bB\cdot\nabla)\bB}{4\pi}+ \rho{\bm a}_{\rm ext,gas} \nonumber\\
    &- \int {\rm d}^3\dvel f_{\rm d}(\dvel){\bm a}_{\rm gas-dust}(\dvel-\gvel) . \label{eq:gmom}
\end{align}
$\bB$ is the magnetic field, and $f_{\rm d}$ is the grain (position-velocity) phase space mass density. Thus, the integral on the right-hand side is merely the opposite of the force imparted to the grains from the gas. Had we included a spectrum of grain sizes (as we intend to in future work), the integral would be over grain size in addition to velocity.

The magnetic field evolves according to the standard induction equation,
\begin{equation}
    \frac{\partial \bB}{\partial t} = \nabla\times(\gvel \times\bB). \label{eq:induction}
\end{equation}
While dust does carry charge, for most astrophysical regimes, the charge it carries is utterly negligible compared to the surrounding gas, and thus we ignore any induction from the dust even in the case where the dust-to-gas mass ratio $\mu > 1$.

Finally, we must also consider the way the total energy of the gas and magnetic field will be affected by the addition of dust. The rate of change of the gas total energy is just,
\begin{equation}
    \frac{{\rm d}\epsilon}{{\rm d}t} = \frac{{\rm d}\epsilon}{{\rm d}t}\bigg|_{\rm MHD} - \int{\rm d}^3\dvel f_{\rm d}(\dvel) {\bm a}_{\rm gas-dust}(\dvel-\gvel)\cdot \dvel, \label{eq:energy}\end{equation}
where $\epsilon$ is the total energy density in the gas and magnetic field (kinetic plus thermal plus magnetic), ${\rm d}\epsilon/{\rm d}t|_{\rm MHD}$ is the usual rate of change in $\epsilon$ in MHD. In essence, we simply need to take into account the work the gas does on the dust, and subtract that from the total energy of the gas. The change in the gas internal energy is implicit in this equation: provided that we know the change in gas kinetic energy (from solving the gas momentum equation) and in magnetic energy (from solving the induction equation), the change in internal energy is given as that which guarantees overall energy conservation.


The ordinary MHD terms here that are present in Equations~\ref{eq:gmom}-\ref{eq:energy} are accounted for by the methods outlined in \cite{teyssier2002cosmological, teyssier2006kinematic} and \citet{fromang2006high}. We use the constrained transport scheme for the magnetic field, guaranteeing that the magnetic field's divergence is zero to machine precision. To evolve fluxes, we the MUSCL-Hancock scheme. We then solve the resulting Riemann problem with the Harten–Lax–van Leer-Discontinuities (HLLD) Riemann solver for gas quantities, and the 2D HLLD Riemann solver for the induction equation. We also employ the MinMod slope limiter to our piecewise linear spatial reconstruction \citep{VanLeer_1979}. Once the MHD updates to the gas and magnetic field have been applied, we then apply our dust updates as described in sections~\ref{sec:lorentz},~\ref{sec:drag}.

\subsection{Units}\label{sec:units}
{We briefly explain here our system of units. Unless otherwise specified, the units used in this paper (often referred to as code units) are fully determined by a unit density (chosen as the mean density most often) $\rho_0$, a unit length $\ell_0$ (most often the box length), and a unit time $\tau_0$ (most often the box length divided by the sound speed). All other quantities can be made dimensionless using these quantities. For example, the gas velocity $\gvel$ and the magnetic field $\bB$ correspond to }
\begin{align}
    \gvel_{\rm code} &\rightarrow \gvel\tau_0/\ell_0 = \gvel/c_{\rm s} \label{eq:velscale},\\
   \bB_{\rm code} &\rightarrow \frac{\tau_0\bB}{\ell_0\sqrt{4\pi\rho_0}} = \frac{\bB}{c_{\rm s}\sqrt{4\pi\rho_0}} \label{eq:bscale},
\end{align}
{where the two equalities on the right hold throughout this paper unless otherwise specified (e.g. if we alter the sound speed so that $c_{\rm s} \neq \ell_0/\tau_0$). In this way, all quantities that we evolve in the code can be expressed in a dimensionless and scale-free way. }
\subsection{Particle-mesh back-reaction}\label{sec:pmbr}
We adapt for our purposes the method outlined in \citet{yang2016integration} termed ``particle-mesh-back-reaction'', which allows for cell-by-cell integration of dust-gas forces and thus avoids the problem of solving a globally coupled system when using implicit methods.\footnote{For the exponential method of \citet{yang2016integration} and implicit methods, the system becomes global rather than local without this method.} We apply this method for both drag and Lorentz forces, for which we use Godunov splitting. In addition to allowing for cell-by-cell integration, this method also helps dust-gas instabilities to grow at the proper rate by allowing nearby dust grains and gas cells to ``talk'' to one another via dust-gas drag. The method is described in detail in \citet{yang2016integration}, but we briefly outline it here. 

Each dust superparticle has a ``cloud'' around it, the interpolation kernel. For us, this is either a uniform cube (as in the cloud-in-cell (CIC) method) or cloud whose density linearly decreases along each dimension (triangular-shape-cloud (TSC)). These clouds overlap with surrounding gas cells. Where a cloud overlaps with a gas cell, treat that overlap as its own ``subcloud''. In each gas cell, compute the new
sub-cloud velocities by first applying Lorentz forces, then applying drag. Add the momentum changes from the sub-clouds back onto the original dust particle, and project its change in momentum back onto the gas using the dust kernel to obtain new gas velocities. 

In the next two sub-sections, we describe our approach to solving for the sub-cloud velocities and an intermediate gas velocity in each cell before adding up the momentum changes to each sub-cloud and projecting the momentum change to each particle back onto the grid.

\subsection{Lorentz force} \label{sec:lorentz}
When examining the dynamics of a uniform dust-gas mixture, the ``average drift velocity'' jumps out as of particular importance and utility. We explore this scenario in Section~\ref{sec:gyrodamp}, where we use the motion of a spatially uniform dust-gas mixture with an initial drift velocity as a test of our algorithm. As well as being a simple test, this problem serves as motivation for the way we have constructed our algorithm. We would like our algorithm to naturally match the expected behavior of this problem, even when the time-step $\Delta t \gg t_{\rm s}, t_{\rm L}$ (the stopping and Larmor times). 

Without back-reaction, the traditional Boris pusher algorithm \citep{boris1970relativistic} combined with an extra drag step is sufficient for following the trajectory of super-particles. However, one can show that when dust grains have finite mass, the Boris algorithm is only stable for $\Delta t \le 2/(\mu \omega_{\rm L})$, with $\mu$ the local dust-to-gas mass ratio and $\omega_{\rm L}$ the local Larmor frequency (assuming uniform grain size). Furthermore, even when this stability criterion is met, grain gyromotions are no longer preserved in the zero-drag limit, instead slowly decaying. We therefore use the insight that, in the presence of back-reaction, it is the \textit{drift velocity} that undergoes gyro-motions, rather than simply the dust (cf. Section~\ref{sec:gyrodamp}).

In this section, we work in a particular cell, letting $j$ index particle sub-clouds (cf. Section~\ref{sec:pmbr}) within that cell. Quantities that have no superscript are taken to be at time $n$ \textit{after} all of the MHD fluxes have been applied. Each sub-cloud initially has the same velocity as its parent-cloud. This cell can be treated as though it is a homogeneous dust-gas mixture with multiple dust ``components'' (as many as there are sub-clouds) superposed on top of one another. 

Define the (mass-weighted) average drift velocity in the cell at time-step $n$,
\begin{align}
    {\bm w}_{\rm s} \equiv \mu^{-1}\sum_j \mu_j {\bm w}_{{\rm s},j} = \mu^{-1}\sum_j \mu_j (\dvel_j - \gvel),
\end{align}
where $\mu$ is the dust-to-gas mass ratio in the cell, $\mu_j$ is the contribution to $\mu$ from sub-cloud $j$, $\gvel$ is the gas velocity. Unless otherwise indicated, all values are taken to be at time $n$ after all of the MHD updates to the gas and magnetic field. 
Then, given a uniform charge-to-mass ratio\footnote{We will incorporate a spectrum of grain sizes and thus charge-to-mass ratio in later work. This requires minor changes to our algorithm.} $\xi$, the Larmor frequency in cell $i$ is $\omega_{\rm L} = \xi B$.
Then a norm-preserving second order accurate update to the average drift velocity is given by the implicit midpoint scheme,
\begin{align}
    {\bm w}_{\rm s}^{n+1} &= {\bm w}_{\rm s} + \Delta t (1+\mu)\omega_{\rm L} \bigg(\frac{{\bm w}_{\rm s}^{n+1} + {\bm w}_{\rm s}}{2}\bigg) \times \hat{\bm b}.
\end{align}
${\bm w}_{\rm s}^{n+1}$ is an estimate of the value of the mass-weighted drift velocity at time $n+1$, $\hat{\bm b}$ is the unit-vector in the direction of the magnetic field, and $\mu$ is the local dust-to-gas mass ratio, both in cell $i$. Solving this to obtain ${\bm w}^{n+1}$, we can then obtain an estimate for the fluid velocity at time $n+1$ as,
\begin{equation}
    \gvel^{n+1} = {\bm V} - \frac{\mu}{1 + \mu}{\bm w}_{\rm s}^{n+1}.
\end{equation}
${\bm V}$ is the center-of-mass velocity in the cell, and $\mu$ is the dust-to-gas mass ratio in the cell. ${\bm V}$ should not change from the application of either Lorentz or drag operators.
Then the new dust sub-cloud velocity estimates are,
\begin{align}
    \dvel^{n+1}_j = \dvel_j + \Delta t\omega_{\rm L}\bigg(\frac{\dvel_j^{n+1} + \dvel_j}{2} - \frac{\gvel^{n+1} + \gvel}{2}\bigg)\times \hat{\bm b}.
\end{align}
This portion of the algorithm is second order accurate, symplectic, and $A-$stable.\footnote{{A method is $A-$stable if when solving the test problem $\dot{y}=ky$, the numerical solution approaches zero as $t\rightarrow\infty$ for all $k\leq0$.}}  This last property is important because in regions where $(1+\mu)\omega_{\rm L}$ becomes large due to a high dust-to-gas mass ratio, the results will still be reasonably physical, albeit inaccurate.  

\subsection{Drag force}\label{sec:drag}
We implement two separate methods for drag: a first order and a second order integrator. The second order integrator is only truly second order for a non-velocity dependent stopping time, otherwise it is simply better than the first order scheme by a constant factor as the time-resolution improves, but will still exhibit first-order convergence.

In this subsection, quantities without a superscript are taken to be at time $n$ after all of the MHD fluxes have been applied, as in the last section, with the exception of the gas velocity and dust velocity, which are taken to be after the Lorentz update (\S~\ref{sec:lorentz}) has been applied.

Define the average stopping rate at time $n$ in each cell as,
\begin{equation}
    \bar{\nu}_{\rm s} \equiv \mu^{-1}\sum_j\mu_j\nu_{{\rm s},j}. \label{eq:mean_nus}
\end{equation}

When we have a constant stopping time, this expression is trivial, but when we have a variable stopping time, using it will introduce an error that degrades our otherwise second order accurate scheme to first order.\footnote{One would also need to implement Strang splitting in the case of a velocity-dependent stopping rate, rather than Godunov, on top of having each solution operator be second order, in order to have the entire scheme be second order in time. We have decided this improvement in accuracy is not worth the computational cost.}

\subsubsection{Variant 1: first order drag update}\label{sec:first_drag}
We advance the average drift velocity to an intermediate value ${\bm w}_{\rm s}^{n+1}$ with the Backward Euler method:
\begin{align}
    {\bm w}_{\rm s}^{n+1} &= {\bm w}_{\rm s} - \Delta t (1+\mu)\bar{\nu}_{\rm s} {\bm w}_{\rm s}^{n+1},\nonumber\\
    &= \frac{{\bm w}_{\rm s}}{1+\Delta t (1+\mu)\bar{\nu}_{\rm s}}.\label{eq:FOdrag_drift}
\end{align}
This value may then be used (together with the cell's center of mass velocity) to compute an intermediate value for the gas velocity. 
\begin{align}
    \gvel^{n+1} &= {\bm V}-\frac{\mu}{1+\mu}{\bm w}_{\rm s}^{n+1}.\label{eq:FOdrag_gas}
\end{align}

Finally, compute the new velocity for the sub-cloud $j$ via,
\begin{align}
    \dvel_j^{n+1} &= \dvel_j -\Delta t \nu_{{\rm s},j} (\dvel_j^{n+1}-\gvel^{n+1}),\nonumber\\
    &= \dvel_j - \frac{\Delta t \nu_{{\rm s},j}(\dvel_j - \gvel^{n+1})}{1 + \Delta t \nu_{{\rm s},j}}
\end{align}
\begin{figure*}
    \centering
    \includegraphics[width=\textwidth]{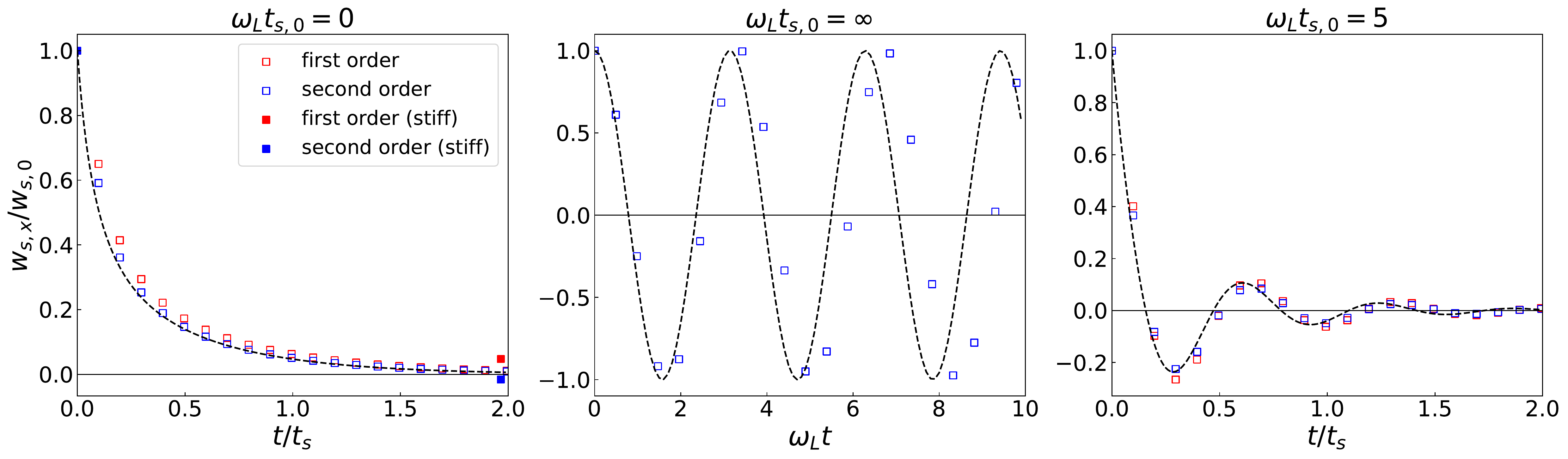}
    \caption{\textbf{Left:} $x$-component of the dust-gas drift velocity $w_{{\rm s},x}$ versus time for our Epstein drag test (velocity-dependent stopping time). The dotted line is the analytic solution from \citet{laibe2012dusty} for the parameters $\mu = 1$, $w_{x,0} = 10c_{\rm s}$. Solid orange points represent our first order implicit algorithm, while solid blue points represent second order (cf. \S ~\ref{sec:drag}). The open points represent the result when a single timestep is taken over two stopping times. \textbf{Middle:} Undamped gyro orbits for large timesteps. $\mu = 1$ here, as in the previous panel. The frequency of the analytic solution is $(1+\mu)\omega_{\rm L}$, so one gyro-period in this plot is $\pi$ in units of $1/\omega_{\rm L}$. A phase error accumulates, but the integrator is symplectic, and the magnitude of the drift velocity remains constant in time.
    \textbf{Right:} Gyro-damping test with the stopping time $\omega_{\rm L} t_{\rm s}(w_{\rm s}\rightarrow 0) = 5$, $\mu = 1$, and $w_{s,0} = 10c_{\rm s}$. The drag law here is Epstein-Baines (Eq.~\ref{eq:kwok}), and the dashed line is an analytic solution (discussed in \S ~\ref{sec:gyrodamp}).}
    \label{fig:gyro_damp}
\end{figure*}
The fact that we use $\nu_{{\rm s},j}$ rather than the mean $\bar{\nu}_{\rm s}$ to update each sub-cloud's velocity restores our accuracy to first order, even though we made a zeroth order error in using $\bar{\nu}_{\rm s}$ to update the mean drift.

\subsubsection{Variant 2: second order drag update}\label{sec:second_drag}
Our second order algorithm is similar. We use a fully implicit midpoint method to obtain an estimate for the mean drift velocity at time $n+1$, and use the midpoint value of the drift to update the gas and dust. For an ordinary differential equation of the form $\dot{y} = f(y)$, the method we employ amounts to having $y^{n+1}, y^{n+1/2}$ simultaneously satisfy,
\begin{align}
    \frac{y^{n+1}-y^{n+1/2}}{\Delta t/2} &= f(y^{n+1}),\\
    \frac{y^{n+1}-y^n}{\Delta t} &= f(y^{n+1/2})
\end{align}
This method is both second-order accurate and $L-$stable (meaning that, in addition to being $A-$stable, the drift velocity will go to zero when the timestep $\Delta t \rightarrow \infty$). It is similar (and for linear equations, equivalent) to \citet{bai2010particle}'s fully implicit trapezoidal method.
We use this method to update the mean drift. The next step is to use the estimate for the mean drift at time $n+1/2$ to estimate the gas velocity at time $n+1/2$. In other words,
\begin{align}
    &{\bm w}_{\rm s}^{n+1/2} = {\bm w}_{\rm s}^{n+1} + \frac{1}{2}\Delta t (1+\mu)\bar{\nu}_{\rm s}{\bm w}_{\rm s}^{n+1},\nonumber\\
    &{\bm w}_{\rm s}^{n+1} ={\bm w}_{\rm s}-\Delta t (1+\mu)\bar{\nu}_{\rm s} {\bm w}_{\rm s}^{n+1/2},\nonumber\\
    &\implies{\bm w}_{\rm s}^{n+1/2} = \frac{1+\Delta t(1+\mu)\bar{\nu}_{\rm s}/2}{1+\Delta t(1+\mu)\bar{\nu}_{\rm s}+\Delta t^2(1+\mu)^2\bar{\nu}_{\rm s}^2/2}{\bm w}_{\rm s}.\label{eq:secondODriftEstimates}
\end{align}
 We then use the midpoint value for the drift velocity to estimate the midpoint gas velocity.The midpoint gas velocity can then be used to determine a midpoint dust velocity that we use to evolve superparticles from time $n$ to $n+1$:
\begin{align}
    \gvel^{n+1/2} &= {\bm V} - \frac{\mu}{1+\mu}{\bm w}_{\rm s}^{n+1/2},\nonumber\\
    \dvel_j^{n+1/2} &= \dvel_j^{n+1} + \frac{1}{2}\Delta t \nu_{{\rm s},j}(\dvel_j^{n+1}-\gvel^{n+1/2}),\nonumber\\
    \dvel^{n+1}_j &= \dvel_j -\Delta t \nu_{{\rm s},j}(\dvel_j^{n+1/2}-\gvel^{n+1/2}) \nonumber\\
    \dvel^{n+1}_j &= \dvel_j - \frac{\Delta t \nu_{{\rm s},j} + \Delta t^2 \nu_{{\rm s},j}^2/2}{1+\Delta t \nu_{{\rm s},j} + \Delta t^2 \nu_{{\rm s},j}^2/2}(\dvel_j - \gvel^{n+1/2}).
\end{align}

We have used our estimate for $\gvel^{n+1/2}$ for all instances of $\gvel$ in the drag force. While for estimating $\dvel_j^{n+1/2}$ this is not second order, the resultant estimate for $\dvel_j^{n+1}$ is indeed second order. 

This second method is subject to minor oscillations that decay rapidly when large time-steps are used, while the first order method we described (\S~\ref{sec:first_drag}) is immune to oscillations. As well, both methods will send gas and dust to the center-of-mass velocity when the time-step becomes much larger than the stopping time. 

The operator-split methods we have presented in this section and the last (\S \ref{sec:lorentz}, \ref{sec:drag}) exhibit what we would consider to be ideal properties for the problem of massive dust grains undergoing damped gyromotions. In the limit where $\omega_{\rm L}\rightarrow0$, the method exhibits exponential decay of the drift velocity, giving reasonable results even in the limit where $\Delta t \gg t_{\rm s}/(1+\mu)$. Where the drag coefficient $\nu_{\rm s} \rightarrow 0$, the method is $A-$stable and is symplectic, with error accumulating only in the phase of the oscillations, but not the magnitude. Even in the limit where $\Delta t \gg t_{\rm s}/(1+\mu) > t_{\rm L}/(1+\mu)$, the behavior of this method is such that dust particles will oscillate within a rapidly decaying envelope, just as they should in the physical solution. Therefore, we do not employ any timestep restrictions based upon the local dust-to-gas mass ratio, for while our results in these regions may be inaccurate when $\Delta t \gg t_{\rm s}/(1+\mu), t_{\rm L}/(1+\mu)$, they are not unphysically so. This contrasts with \citet{bai2010particle}'s method of artificially increasing the stopping time $t_{\rm s}$ in regions where $\mu$ becomes large.

\section{Results} \label{sec:results}
The following sections detail the results of various numerical tests we have used to validate our code. These tests also build physical intuition for some of the processes by which grains may be accelerated in our simulations. 

First, we will analyze the results of the simplest possible test of our new methods, damped dust-gas gyromotion in a uniform medium (\S~\ref{sec:gyrodamp}). This test serves to isolate the dust dynamics and back-reaction implementation of our code. We then examine two different dust acceleration scenarios, a shock propagating through a uniform dust-gas medium (\S~\ref{sec:dusty_shocks}) and Alfv{\'e}n wave gyro-resonance (\S~\ref{sec:dusty_alfven}). In Section~\ref{sec:orszag-tang} we present the results of dusty Orszag-Tang vortex \citep{orszag1979small} simulations and illustrate that the same features that appear in the shock test also appear in a more complex, multi-dimensional problem. Section~\ref{sec:rdi} shows the results of a simulation of the resonant drag instability (RDI) chosen to be as close as possible to that studied in SHS19. As the RDI does not grow without dust back-reaction, this test validates our back-reaction implementation.

The final test we perform, decaying turbulence, also serves to lead into our follow-up paper \citep[][in prep.]{MoseleyInPrep} that will study driven turbulence with a range of grain sizes. 
\subsection{Damped gyro-motions}\label{sec:gyrodamp}
One simple test problem that tests drag, Lorentz forces, and back-reaction is that of dust uniformly gyro-orbiting gas. Dust and gas are here spatially homogeneous. This lets us isolate the performance of the dust solution operators that we have described in the previous sections~\ref{sec:lorentz}, \ref{sec:drag}. As well, the magnetic field follows exactly the $z$-axis and is also spatially uniform. 

In this simplified scenario, the Equations \ref{eq:dustvel} and \ref{eq:gmom} can be rewritten with complex quantities instead of vector quantities. That is to say, we define complex quantities where, e.g.
\begin{equation}
    \tilde{\dvel}\equiv v_x + i v_y.
\end{equation}
We use tildes to denote the complex versions of vector quantities. This notation enables us to compactly express both the dynamical equations and solutions without the need for vector or matrix algebra. It can be regarded as a purely notational choice.

We can then solve the following equations for the dust and gas velocities:
\begin{align}
    \frac{{\rm d}\tilde{\dvel}}{{\rm d}t} &= -(\nu_{\rm s}+i\omega_{\rm L})\tilde{\drift}\\
    \frac{{\rm d}\tilde{\gvel}}{{\rm d}t} &= \mu(\nu_{\rm s} + i\omega_{\rm L})\tilde{\drift}.
\end{align}
Without loss of generality, let the center-of-mass velocity be zero. Then we only need to solve for the drift velocity
\begin{align}
    \frac{{\rm d}\tilde{\drift}}{{\rm d}t} &= -(1+\mu)(\nu_{\rm s}+i\omega_{\rm L})\tilde{\drift},
\end{align}
which admits simple solutions. For the case where $\nu_{\rm s}$ is constant,
\begin{equation}
    \tilde{\drift} = \tilde{{\bm w}}_{{\rm s},0}\exp\big[-(1+\mu)(\nu_{\rm s} + i \omega_{\rm L})t\big].
\end{equation}
For Epstein-Baines drag (Eq.~\ref{eq:kwok}), we have:
\begin{align}
    &\tilde{\drift} = \frac{c_{\rm s}}{\sqrt{\eta}}\frac{\tilde{\bm w}_{{\rm s},0}}{|\tilde{\bm w}_{{\rm s},0}|}\exp(-i(1+\mu)\omega_{\rm L} t)\times \nonumber\\
    &\sqrt{\Bigg(\frac
    {\sinh{\big((1+\mu)\nu_{{\rm s},0} t\big)}+\sqrt{1+\eta|\tilde{\bm w}_{{\rm s},0}|^2/ c_{\rm s}^2}\cosh{\big((1+\mu)\nu_{{\rm s},0} t\big)}}
    {\cosh{\big((1+\mu)\nu_{{\rm s},0} t\big)}+\sqrt{1+\eta|\tilde{\bm w}_{{\rm s},0}|^2/ c_{\rm s}^2}\sinh{\big((1+\mu)\nu_{{\rm s},0} t\big)}}
    \Bigg)^2-1},\label{eq:ncgyrodamp}\\
    &\eta \equiv \frac{9\pi\gamma}{128}, \nonumber\\
    &\nu_{{\rm s},0} \equiv \sqrt{\frac{8}{\pi\gamma}}\frac{\rho c_{\rm s}}{\rho_{\rm d}^{\rm i} \agr}.\nonumber
\end{align}
Here, $\tilde{\bm w}_{{\rm s},0}$ is the initial drift velocity between gas and dust. The non-oscillatory drag factor above comes from \citet{laibe2012dusty}, while the oscillatory exponential term trivially follows. We can see that as the dust-to-gas ratio increases, both the effective stopping rate $(1+\mu)\nu_{\rm s}$ and the effective gyro-frequency $(1+\mu)\omega_{\rm L}$ increase. This effect only becomes significant for $\mu\gtrsim 1$, which may happen in small isolated portions of turbulent dusty clouds \citep{bai2010particle, hopkins2016fundamentally, lee2017dynamics, moseley2019non, Beitia+Gomez+Vallejo_2021}. 

\begin{figure*}
    \centering
    \includegraphics[width=\textwidth]{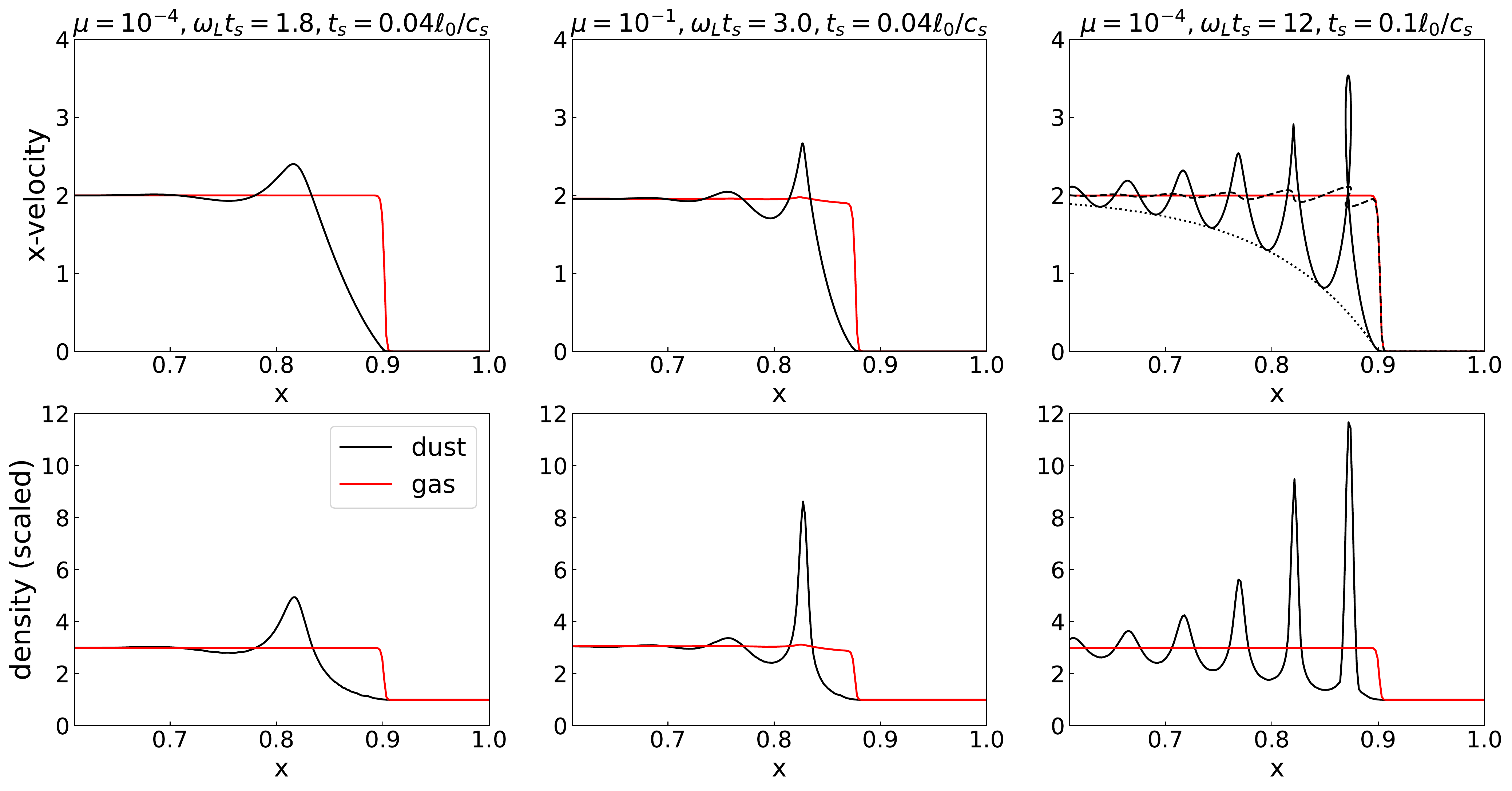}
    \caption{$512^3$ simulations of isothermal dusty shock tubes with single-size dust grains for varying dust-to-gas mass ratios $\mu$, magnetization $\omega_{\rm L} t_{\rm s}$, and stopping time $t_{\rm s}$ shown at time $t=0.3 \ell_0/c_{\rm s}$. Further description of the setup can be found in Section~\ref{sec:dusty_shocks}. {\bf Top row:} $x$-velocity versus position $x$ for dust (black) and gas (red). The panel in the top right has two additional curves: the guiding-center velocity of the grains (dashed, black) and the $\mu=0$ analytic solution for neutral dust grains with the same stopping time (dotted, black). {\bf Bottom row:} The (scaled) dust and gas densities versus position for the same simulations. Dust densities are computed as specified in Section~\ref{sec:dusty_shocks} and then scaled by a factor of $\mu^{-1}$ for visibility. In the middle column where $\mu=0.1$, we can clearly see the effect of the dust back-reaction on the shock. The velocity of the gas ticks up slightly where the highest dust-velocity peak is. This same behavior is mirrored in the density. These dust-gas waves are analogous to ion-cyclotron waves. A further effect of back-reaction is to slow the propagation of the shock.}
    \label{fig:dusty_shocks}
\end{figure*}

\begin{figure}
    \centering
    \includegraphics[width=0.85\columnwidth]{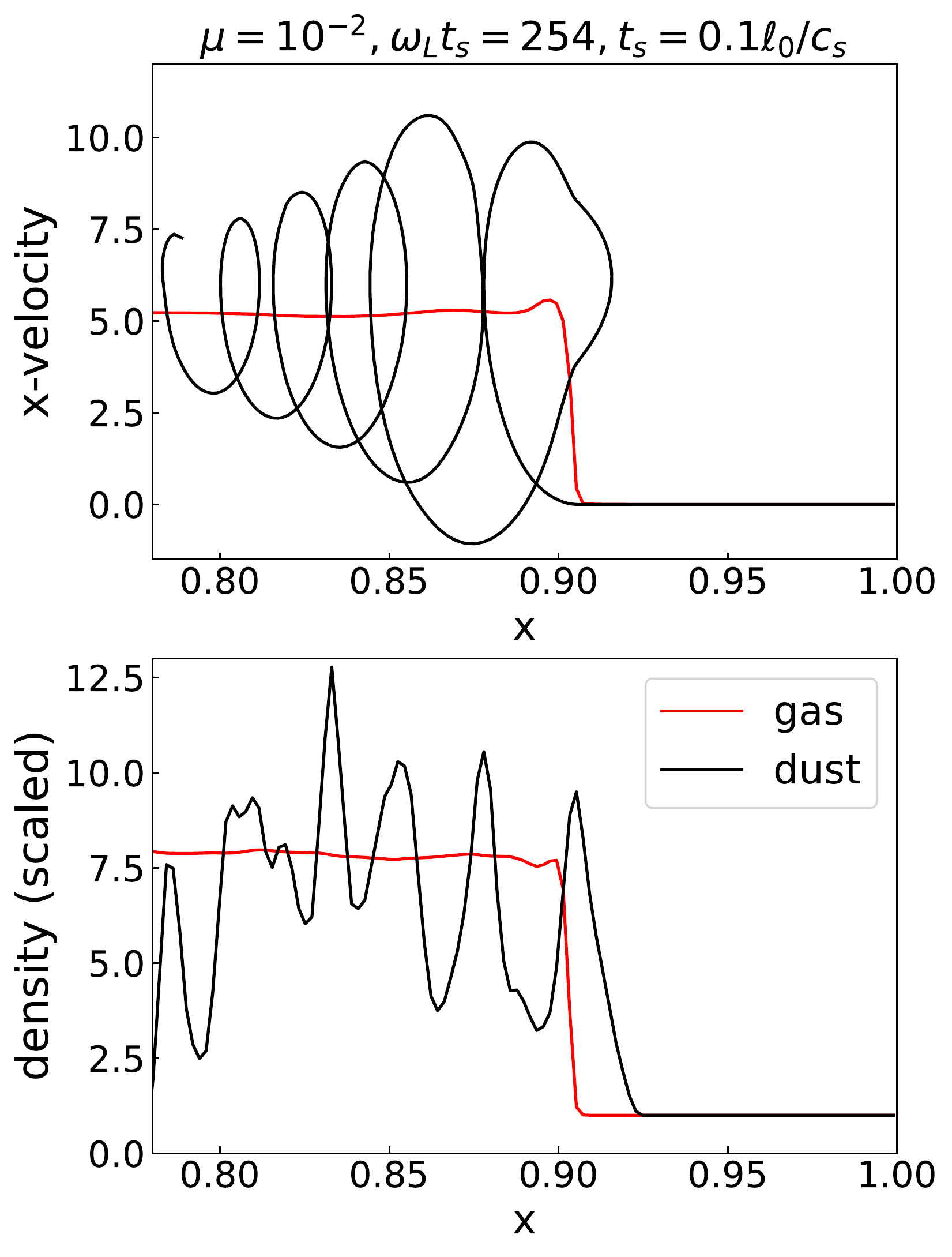}
    \caption{{The figure above is similar to that in fig.~\ref{fig:dusty_shocks}, but the sonic Mach number is 100 and the Alfv{\'e}n Mach number is 6. As well, grains are more strongly magnetized, so that $\omega_{\rm L} t_{\rm s} \approx 254$. These parameters make this shock more comparable to those shown in \citet{guillet2009shocks}. The snapshot is taken at time $t=15 \ell_0/c_{\rm s}$. Further description of the setup can be found in Section~\ref{sec:dusty_shocks}. {\bf Top row:} $x$-velocity versus position $x$ for dust (black) and gas (red). Fermi acceleration is visible, with grains exiting and reentering the shock face and reaching a higher peak velocity upon reentry. Additionally, the effect of the dust back-reaction is more pronounced with this case than with the shocks in fig.~\ref{fig:dusty_shocks}.  {\bf Bottom row:} The (scaled) dust and gas densities versus position for the same simulation. Dust densities are computed as specified in Section~\ref{sec:dusty_shocks} and then scaled by a factor of $\mu^{-1}$ for visibility. The density varies wildly because of the many orbit-crossings visible in the top panel. This also demonstrates that the dust to gas mass ratio can vary quite wildly when grains are all of a similar size, even in a smooth and simple initial setup with no turbulence at all. }}
    \label{fig:mach6}
\end{figure}

We compare the results from our code against the analytic solution given by Equation~\ref{eq:ncgyrodamp}. The results of this are shown in Figure~\ref{fig:gyro_damp}. We intentionally use very low time-resolution to highlight stability and numerical errors. We show $\omega_{\rm L} t_{{\rm s},0} = 0,5,\infty$ for both our first and second order drag implementations (described in Section~\ref{sec:drag}). The initial drift velocity is ${\bm w}_{{\rm s},0} = 10c_{\rm s}\hat{\bm x}$ for all simulations, and all simulations use the Epstein-Baines drag law (Eq.~\ref{eq:kwok}).

\subsection{Dusty shocks}\label{sec:dusty_shocks}
In this subsection, we explore dusty shocks. The simulations we present in this section serve both to help us understand the origin of the dust distribution in phase-space that we see in Section~\ref{sec:decayturb} as well as to validate our methods. Shocks have been considered as a site of dust-processing, eroding, vaporizing and shattering grains depending on the particular shock parameters \citep{melandso1995theory, schilke1997sio, guillet2007shocks, guillet2009shocks, guillet2011shocks, anderl2013shocks}. As well, grains may modify shock structure through momentum feedback, varying the ionization fraction, and dissipation of energy via infrared cooling, friction, dust vaporization, and stochastic charging on small grains. 

Of these past studies, the most comparable to the simulations we present here is \citet{guillet2009shocks}, wherein the authors studied the impact of J-shocks on dust destruction, as well as the trajectories and velocities of grains. The scenario considered by \citet{guillet2009shocks} differs somewhat from that which we simulate here. They allow for heating and cooling throughout the shocks, resulting in temperatures that span a range from 10\K~through 10$^5$\K. They model the entirety of the shock structure, beginning at the upstream equilibrium all the way through to the establishment of new equilibrium temperatures, densities, and magnetic field strengths. In our case shocks are isothermal, with sonic and Alfv{\'e}n Mach numbers of 3. While this makes our Alfv{\'e}n Mach number similar to that in \citet{guillet2009shocks} ($\mathcal{M}_{\rm A}\approx 1.3-2.4$, depending on the simulation), the sonic Mach numbers studied in \citet{guillet2009shocks} are well in excess of 100. Regardless, the grain drift velocities in units of the Alfv{\'e}n speed in their simulations and ours should be comparable. More specifically, grain velocities should reach a maximum that is of order the shock velocity in the post-shock region, barring higher Alfv{\'e}n Mach numbers that may result in Fermi acceleration. Ignoring collisions, this makes our simulations and theirs somewhat comparable.

Each shock simulation we present here has the same initial conditions and boundary conditions. As we have only implemented a 3D version of our code at this time, these tests are performed in three dimensions with the shock set up to propagate exactly along one of the three cardinal directions, orthogonal to the magnetic field.
\begin{figure*}
    \centering
    \includegraphics[width=\textwidth]{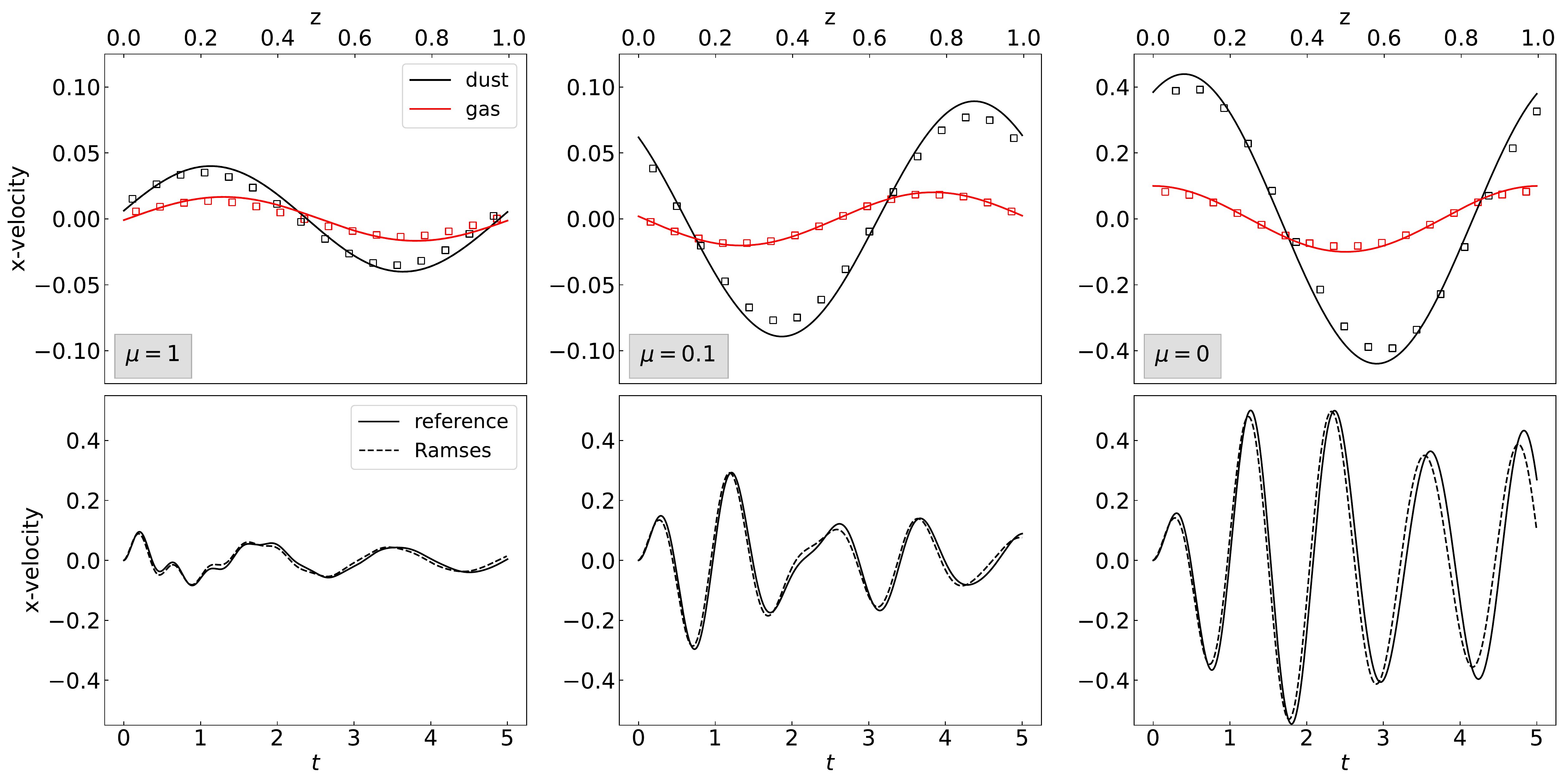}
    \caption{\textbf{Top row:}
    Circularly polarized, dusty Alfv{\'e}n wave test at $t=5$ for varying dust-to-gas mass ratios $\mu$ at a resolution of $N=16^3$ dust particles and gas cells, {where $\Omega_{\rm AW}$ is the Alfv{\'e}n wave frequency. In these units, the Alfv{\'e}n wave period is 1.} The gas mass is the same in each simulation, while the dust mass varies. Solid lines show a high-accuracy reference solution computed as described in Section~\ref{sec:dusty_alfven}, while squares show results from the RAMSES simulations. Details on the numerical setup can also be found in Section~\ref{sec:dusty_alfven}. The dust charge-to-mass ratio is chosen to be exactly resonant with the Alfv{\'e}n wave. The drag coefficient is constant, and $-\larmorfreq t_{\rm s} = 10$, the amplitude $\delta B = \delta u = 0.1$, and $-\larmorfreq = \Omega_{\rm AW} = 2\pi v_{\rm A}/\ell_0$. Dust is initially moving uniformly along the propagation direction (see Section~\ref{sec:dusty_alfven}). \textbf{Bottom row:} The x-velocity of a single dust grain versus time in our low-resolution simulation with RAMSES vs as computed with the high-accuracy reference. To see alternate gyroperiods, see figure~\ref{fig:dusty_alfven_varom}.}
    \label{fig:dusty_alfven_varmu}
\end{figure*}

\begin{figure*}
    \centering
    \includegraphics[width=\textwidth]{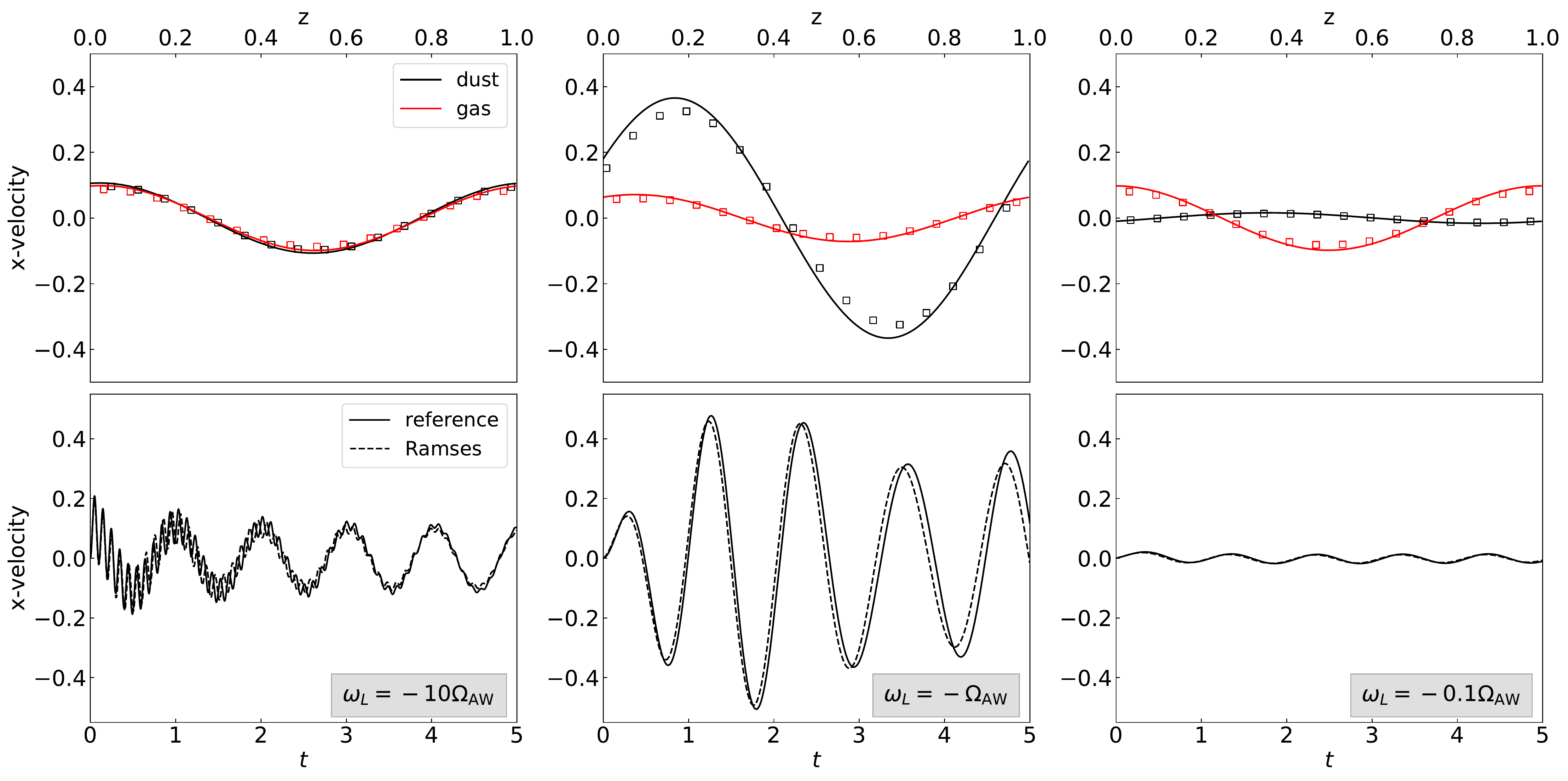}
    \caption{The same as Figure~\ref{fig:dusty_alfven_varmu}, but rather than varying the dust-to-gas mass ratio $\mu$, we fix $\mu = 0.01$ and $\Omega_{\rm AW}=2\pi v_A/\ell_0$ while varying the Larmor frequency of grains $\omega_{\rm L}$. The stopping time is also fixed at $t_{\rm s} = 0.1 \ell_0/c_{\rm s}$. In the leftmost column, we see that a high Larmor frequency means the grains efficiently couple to the Alfv{\'e}n wave, not notably changing the wave's phase  nor amplitude. For perfectly resonant dust grains, we see that grains reach a maximum velocity just slightly more than four times the original wave amplitude while also shifting the phase of the original Alfv{\'e}n wave through dispersion introduced by the grains. In the rightmost column, where grains have just one tenth of the resonant frequency as their Larmor frequency, the grains hardly respond to the waves.
    }
    \label{fig:dusty_alfven_varom}
\end{figure*}
In Figure~\ref{fig:dusty_shocks}, a Mach 3 isothermal shock propagates from left to right along the $x$-axis into an at-rest medium. For the simulation boundaries orthogonal to the shock's direction of propagation, we use periodic boundary conditions. Each of our coordinates, $x,y,z$ go from zero to one. At $x = 0$, we set the boundary such that we meet the shock-jump conditions\footnote{This inflow boundary can lead to minor wall-heating, but does not impact the downstream solution far away from the boundary.}  (ignoring dust):
\begin{align}
    \bB(x=0) &= 3\hat{\bm z},\nonumber\\
    \rho(x=0)&= 3,\nonumber\\
    \gvel(x=0) &= 2\hat{\bm x}.
\end{align}
Right of the shock (in the at-rest medium), we have:
\begin{align}
    \bB(x>0) &= \hat{\bm z},\nonumber\\
    \rho(x>0)&= 1,\nonumber\\
    \gvel(x>0)&= {\bm 0}.
\end{align}
Right of the shock, the dust velocity is zero. Left of the shock, there is initially no dust (we do not currently implement the creation of dust grains that would be necessary for this). For the boundary at $x=1$, we use an outflow boundary condition. 

{In Figure~\ref{fig:mach6}, we show a shock whose sonic Mach number is 100 and Alfv{\'e}n Mach number is 6, which is more easily compared to the results in \citet{guillet2009shocks}. The Alfv{\'e}n Mach number here is high enough that we can observe Fermi acceleration. The Alfv{\'e}n Mach number here is more important for the grain orbital behavior than the sonic Mach number, for it is the Alfv{\'e}n Mach number which determines whether or not grains cross the shock front more than once. }
{The initial conditions right of the shock shock are the same as for the shocks described above, but we impose a boundary condition on the left such that,}
\begin{align}
    \bB(x=0) &= 7.99661\hat{\bm z},\nonumber\\
    \rho(x=0)&= 7.99661,\nonumber\\
    \gvel(x=0) &= 5.24968\hat{\bm x},\nonumber\\
    c_{\rm s} &= 0.06 \,({\rm everywhere}).
\end{align}

In Figure~\ref{fig:dusty_shocks}, we show the results of shock simulations with this setup run at varying dust-to-gas mass ratio $\mu$, magnetization $\omega_{\rm L} t_{\rm s}$, and stopping time $t_{\rm s}$. For all of these shocks, we use a constant drag law, CIC interpolation, and our second order integrator (cf. Section~\ref{sec:second_drag}). At low values of magnetization, the dust velocity remains single-valued (no orbit-crossings). At high values of magnetization $\omega_{\rm L} t_{\rm s}$, orbit-crossing occurs, introducing two discontinuities in the density, each occurring when the dust velocity transitions from being single-valued to multi-valued and vice-versa. This is because there is an abrupt transition from the grain velocities being single-valued with position to triple-valued. Thus, the density is higher in the triple-valued region than in the single-valued region, simply because there are more grains at the same position coordinate.

High levels of magnetization also more effectively couple dust to gas. This can be seen in the guiding center velocity, which when accounting for the constant drag law and the geometry of this problem is given by,
\begin{align}
    V_x = u_x -\frac{w_{{\rm s},y}}{\omega_{\rm L} t_{\rm s}},\\
    V_y = u_y + \frac{w_{{\rm s},x}}{\omega_{\rm L} t_{\rm s}}.
\end{align}
$V_x$ can be seen as the dashed line in the upper right panel of Figure~\ref{fig:dusty_shocks} for our highest magnetization shock simulation. In the same panel, there is also a dotted line representing the analytic solution for grain velocity versus position with no gyro-motion. It is worth noting that the actual drift velocity does not decay any faster or slower in the case of charged grains vs neutral grains, even though in an orbit-averaged sense the charged grains are more effectively coupled to the gas. 

When the dust-to-gas mass ratio is increased so that dust back-reaction is non-negligible, we see in the center column of Figure~\ref{fig:dusty_shocks} that dust launches low-amplitude waves analogous to ion-cyclotron waves in ordinary plasma. One might thus call these `dust-cyclotron' waves. These are visible as a small perturbation on the otherwise uniform shape of the shock. The highest peaks of these waves correspond to the maximum peaks of the dust velocity and density. These wave modes were also formerly recognized in the `gyroresonant' modes explored in \citet{hopkins2018ubiquitous}, although in that work, they were explored in the context of resonant drag instabilities (RDIs) (cf. Section~\ref{sec:rdi}).

It's worth clarifying here how we compute dust density throughout this work, unless otherwise specified. First, in post-processing, the mass of the grains is deposited onto the grid using the CIC (cubic) kernel. Then, the density is convolved with a small gaussian kernel with a standard deviation of one cell. 

One thing to note is that this simplified, low Alfv{\'e}n Mach number shock setup can never accelerate grains to any higher drift velocity than the initial jump in velocity across the shock. When $\mu \rightarrow 0$, for single grains, this problem is physically identical to that of uniform grain motion in Section~\ref{sec:gyrodamp}: at some specified time, an initial drift velocity $\Delta {\bm w}_{\rm s}$ is set, and the drift velocity then gyro-rotates and damps away, never reaching a value any higher than its initial value. Low Alfv{\'e}n Mach number shocks of this kind thus do not accelerate charged grains to any higher drift velocities than uncharged grains. This is, however, untrue for high Alfv{\'e}n Mach number shocks for which well-magnetized grains can enter the pre-shock medium at least once, undergoing Fermi acceleration.

\subsection{Dusty Alfv{\'e}n wave}\label{sec:dusty_alfven}

\begin{figure*}
    \centering
    \includegraphics[width=\textwidth]{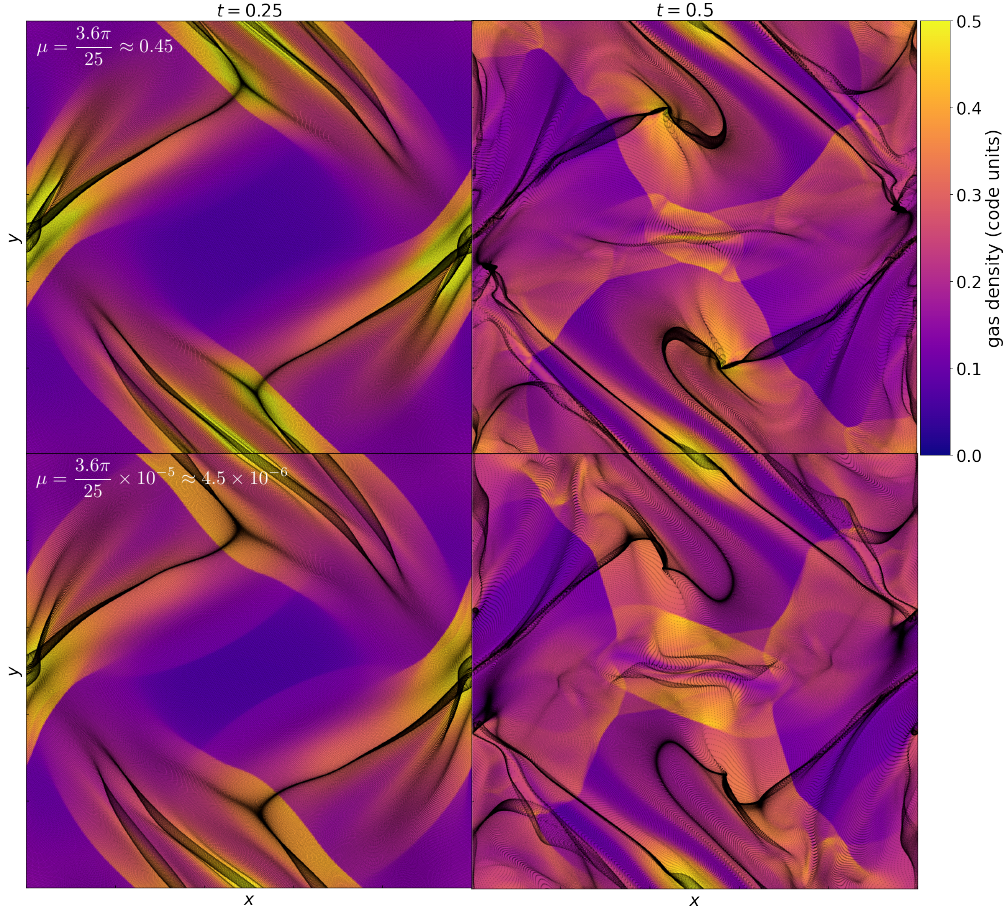}
    \caption{Slices of the (3D) dusty Orszag-Tang vortex at times $t=0.25$ (left) and $t=0.5$ (right) for dust-to-gas mass ratios $ \mu =3.6\pi/25 \approx 0.45$ (top) and $3.6\pi\times10^{-5}/25 \approx 4.5\times10^{-6}$ (bottom). Both simulations have a resolution of $N_{\rm dust} = N_{\rm gas}=512^3$. Dark points represent the positions of dust grains, while the underlying color scheme shows gas density. The dust grains have a constant stopping time of $t_{\rm s} = 0.1$ and a dimensionless charge-to-mass ratio $\xi = 100$.}
    \label{fig:dustyot}
\end{figure*}
\begin{figure}
    \centering
    \includegraphics[width=\columnwidth]{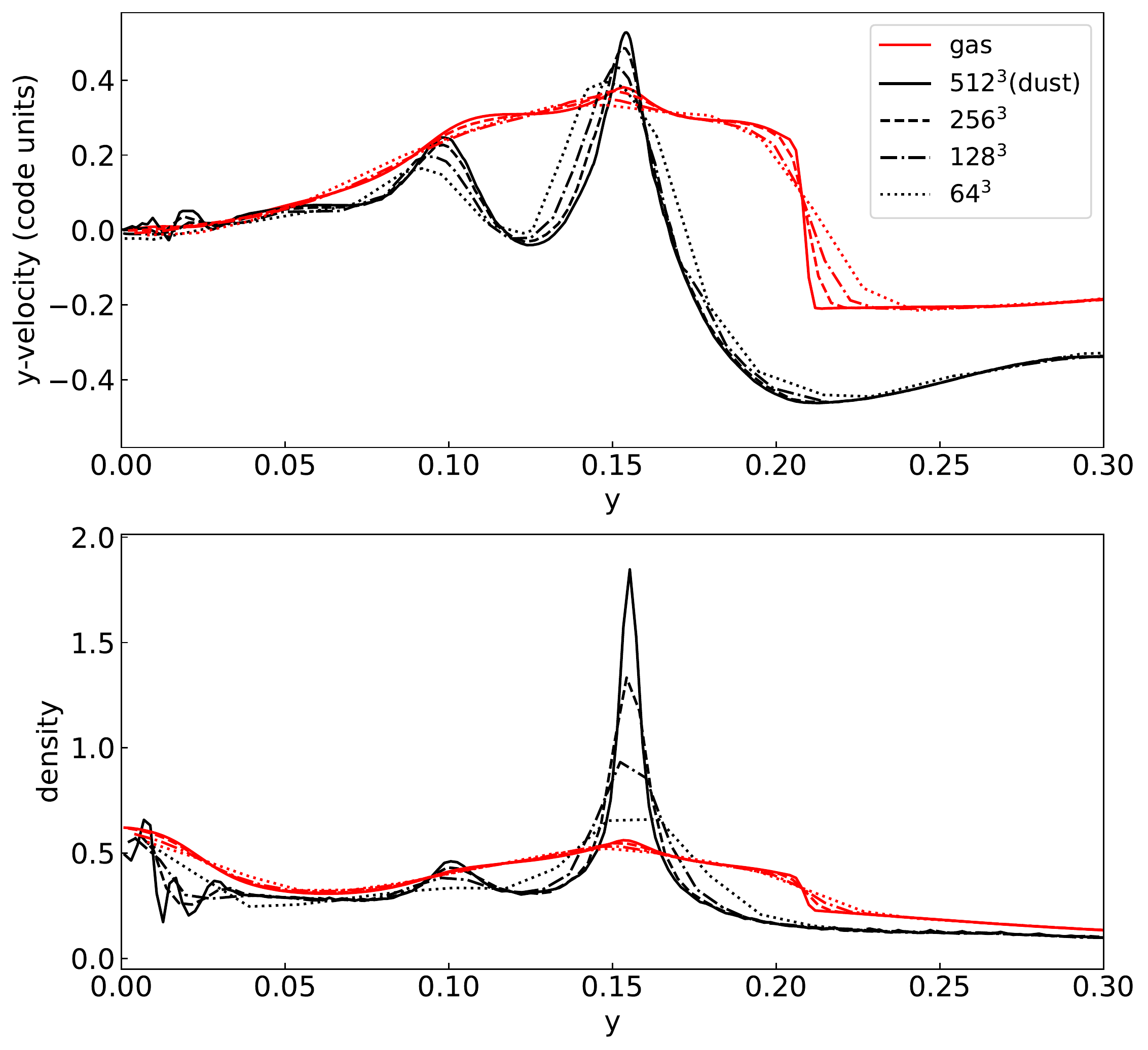}
    \caption{\textbf{Top:} A slice of the dust velocity (black) and gas velocity (red) for the dusty Orszag-Tang problem at $t=0.25$. The slice is taken at $x=0.5$. We only show the solution up to $y=0.3$ because it is virtually uniform beyond this point until the other side. These profiles are computed as described in Section~\ref{sec:orszag-tang}. To see the entire domain at a resolution of $N_{\rm dust} = N_{\rm gas}=512^3$, see Figure~\ref{fig:dustyot}.
    \textbf{Bottom:} The same as above, but dust density and the gas density instead of velocity. Density is in code units, and the initial (i.e. mean) dust-to-gas mass ratio is  $\mu= 3.6\pi/25 \approx 0.45$. }
    \label{fig:otslice}
\end{figure}
We next explore the problem of a circularly polarized dusty Alfv{\'e}n wave. In code units, the initial conditions are:
\begin{align}
    \gvel(t=0) &= \bB(t=0) &= u_0\cos(2\pi z)\hat{\bm x} + u_0\sin(2\pi z)\hat{\bm y} + \hat{\bm z}\\
    \dvel(t=0)&=\hat{\bm z}
\end{align}
The equation of state is isothermal and $P = \rho = 1$.  We let $u_0 = 0.1$ for this problem, and the box size is $\ell_0 = 1$. We choose the dust gyro-frequency to be exactly resonant with the Alfv{\'e}n wave frequency, $\omega_{\rm L} = -2\pi$, and the drag such that $|\omega_{\rm L}| t_{\rm s} = 10$, so that grains are reasonably well magnetized. The drag coefficient $\nu_{\rm s}$ is constant, and we use our second-order integrator (cf. \S~\ref{sec:second_drag}). We also use TSC interpolation. 
The symmetry of this problem should be maintained for all times $t>0$, so that rotation through an angle $2 \pi \Delta z$ and translation by $\Delta z$ are exactly equivalent. That is to say, the solutions remain perfectly circularly polarized in all variables with the phases changing with time along with the amplitudes. This fact can be leveraged to write a numerical integrator for the system that operates only at one point, after which the solution can be ``propagated'' to the rest of the interval trivially. This is the method by which we obtain the high-accuracy reference solutions shown in Figures~\ref{fig:dusty_alfven_varmu} and \ref{fig:dusty_alfven_varom}, using $10^6$ time-steps between $t=0$ and 5.

Figures~\ref{fig:dusty_alfven_varmu} and \ref{fig:dusty_alfven_varom} show comparisons after 5 periods between the results obtained with $N=16^3$ dust particle and gas cell resolution simulations in RAMSES and the aforementioned reference solutions. We use varying dust masses, with $\mu = 0,0.01,0.1,1$. Even this low resolution, the wave is fairly accurately modeled by RAMSES. At resolutions of $N=32^3$ and above, the results of RAMSES become almost indistinguishable from the reference.

We have tried a range of amplitudes and drag coefficients (not shown). We find there are multiple mechanisms for saturation of the dust grain amplitude. One is that if grains move near the Alfv{\'e}n speed, non-linear terms generate significant motions in $z$ (already present for even an amplitude of $0.1 v_A$ in figs.~\ref{fig:dusty_alfven_varmu},~\ref{fig:dusty_alfven_varom}), causing a Doppler shift of the Alfv{\'e}n frequency and breaking the resonance. Another is that if grains have sufficient mass, they cannot have more energy and momentum dumped into them than is available from the wave. The final mechanism is that minute motions in $z$ produce a phase error that accumulates so that dust grains eventually fall into anti-phase with the wave and lose speed as a result. This is not the same mechanism as the Doppler shift causing resonance breaking: the motions may be small enough that the frequency of the wave in the frame of the dust should still be strongly resonant. An implication of these mechanisms is that gyro-resonance with a single, pure, circularly polarized Alfv{\'e}n mode cannot accelerate grains to faster than the Alfv{\'e}n speed.

One caveat of our method arises in the high-drag regime. \citet{laibe2012dusty} showed that sound waves are over-damped when the stopping-length $c_{\rm s} t_{\rm s}$ is under-resolved. We find this is similarly true for Alfv{\'e}n waves for sufficiently short stopping times at high dust-to-gas mass ratio; the waves are damped more than they physically should be, though the problem is not as severe as with sound waves in \citet{laibe2012dusty}. Further, at low dust-to-gas mass ratio (e.g. $\mu = 0.01$), this issue does not seem to manifest in any significant way, even at $t_{\rm s} = 10^{-4}\ell_0/c_{\rm s}$, shorter than the time-step for the low-resolution simulations presented here.
\begin{figure*}
    \centering
    \includegraphics[width=\textwidth]{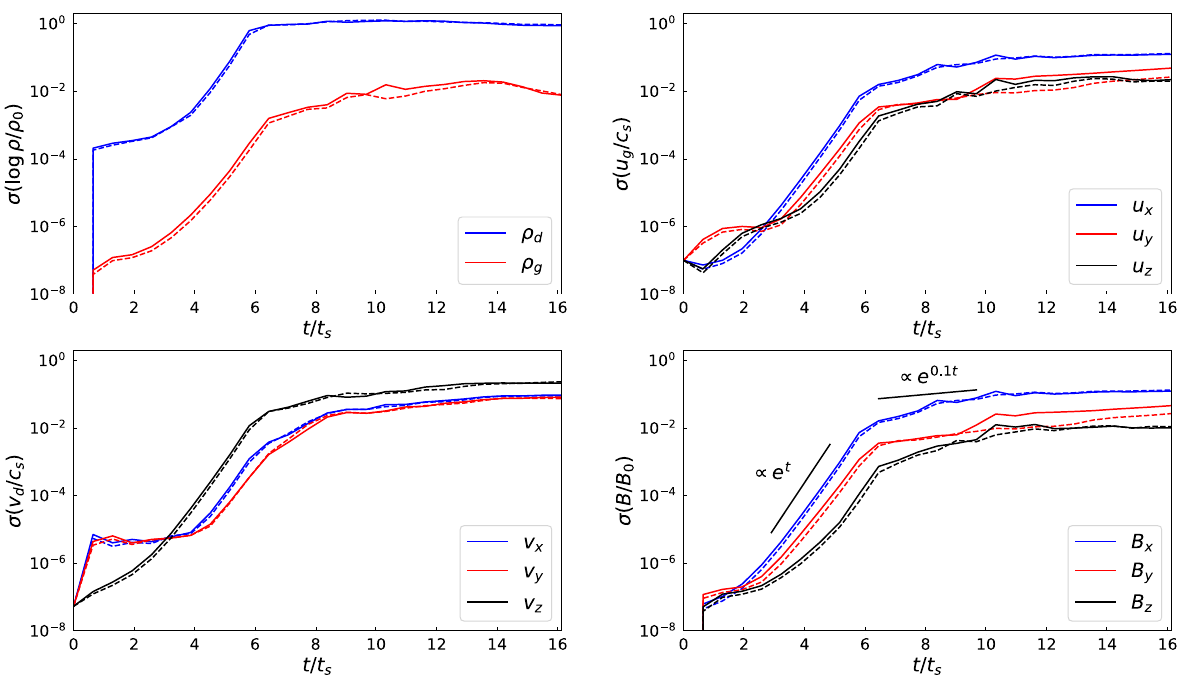}
    \caption{Growth of the magnetic RDI with parameters identical to those in SHS19 for CIC (dashed) and TSC (solid) interpolation. When comparing to Figure 5 in SHS19, we see that the properties of the saturated state seen in RAMSES are almost identical to those seen in GIZMO. As well, the growth rate we observe is the same. The lines $\exp(t)$ and $\exp(0.1t)$ here are chosen for comparison to SHS19, who included the same lines. The time $t$ in the exponent is in code units, i.e. units of $\ell_0/c_{\rm s}$. For further comparison, see a 3D visualization of the instability in Figure~\ref{fig:rdi_cubes}.
    }
    \label{fig:rdi_growth}
\end{figure*}
\begin{figure}
    \centering
    \includegraphics[width=0.95\columnwidth]{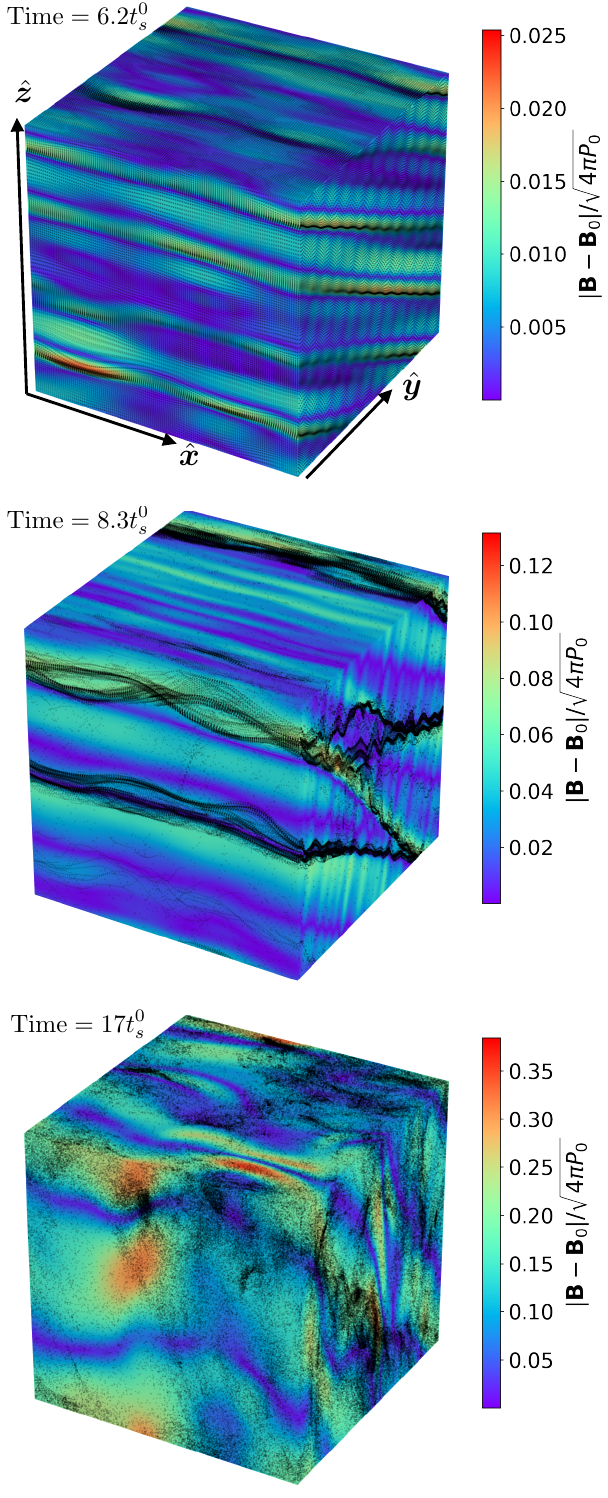}
    \caption{3D visualization of the magnetic RDI with parameters identical to the simulation from SHS19. We show snapshots from times corresponding to the linear regime (top), early non-linear regime (middle) and saturated state (bottom). Colors represent the strength of the magnetic field fluctuations along each face of the cubes, while dark points show the positions of dust grains within thin slices on the edges of the cubes. This simulation is strikingly similar to the SHS19 simulation, showing the same qualitative evolution, growth rate (cf. fig.~\ref{fig:rdi_growth}), resonant angles, and saturation amplitudes.
    }
    \label{fig:rdi_cubes}
\end{figure}
\subsection{Dusty Orszag-Tang vortex}\label{sec:orszag-tang}
We have also simulated a ``dusty'' version of the classic Orszag-Tang vortex problem \citep{orszag1979small}. The most important difference between the ordinary (compressible) version of the problem is that we add on an extra dust component. This increases the amount of mass in the system by a factor of $1+\mu$. Dust superparticles are initialized at the center of each cell and given the same velocity as that cell, i.e. dust initially moves with gas. The initial conditions for the problem are,
\begin{align}
    \dvel(0)&=\gvel(0) = -\sin(2\pi y)\hat{\bm x} + \sin(2\pi x)\hat{\bm y}\\
    \bB(0) &= \frac{1}{\sqrt{4\pi}}\left(-\sin(2\pi y)\hat{\bm x}+\sin(4\pi x)\hat{\bm y}\right),\\
    \rho(0) &= \frac{25}{36\pi},\\
    P(0) &= \frac{5}{12 \pi},\\
    \gamma &= 5/3,\\
    \varepsilon &= 0.1,\\
    \xi &= 100,\\
    \nu_{\rm s} &= \nu_0 = 10,\\
    \rho_{\rm d} &= 10^{-1},10^{-6}
\end{align}

The dust density $\rho_{\rm d}$ being $10^{-1}$ or $10^{-6}$, depending on the simulation (cf. fig.~\ref{fig:dustyot}) gives us dust-to-gas mass ratios of $\mu = 3.6\pi/25\approx 0.45$ and $3.6 \times10^{-5}/25\approx 4.5\times 10^{-6}$.

Again, as our code currently only allows for 3D simulations, we run a 3D version of this 2D problem. We have run this problem for resolutions ranging from $64^3$ to $512^3$ dust particles and gas cells. Our resolution study was done for a mean dust-to-gas mass ratio of $\mu \approx0.45$). A single $512^3$ simulation with $\mu = 3.6\times10^{-5}/25 \approx 4.5 \times 10^{-6}$ ($\rho_{\rm d} = 10^{-6}$) was run to illustrate the effect that massive dust has on the problem. The grains for both the high dusty density $\rho_{\rm d}= 10^{-1}$ and low density $10^{-6}$ simulations use a constant dust stopping time with $t_{\rm s} = 0.1$ (in code units). We use our second order integrator (\S~\ref{sec:second_drag}) and TSC interpolation for these tests, so that the simulations presented here should be second order in both space and time.

Thin slices of the results of the $512^3$ simulations are shown in Figure~\ref{fig:dustyot}. It is worth noting that this is the only simulation we present in this paper where dust grains are allowed to heat the gas (cf. Eq.~\ref{eq:energy}), for all of our other simulations are isothermal. Thus, some of the differences seen in the gas density distribution between the $\rho_{\rm d} = 10^{-1}$ and $10^{-6}$ simulations are due to heating.\footnote{We have checked this explicitly by running simulations with and without heating.}

Comparing the $\rho_{\rm d} = 10^{-6}$ simulation to the $\rho_{\rm d} = 10^{-1}$ simulation, we see that the presence of massive dust slows the propagation of shocks noticeably. This can be seen through the location of shocks in each of the snapshots: the shocks have proceeded further for $\mu=10^{-6}$ than $10^{-1}$.

Visible in both simulations is the presence of dust density oscillations exactly analogous to those seen in our dusty shock simulations (\S~\ref{sec:dusty_shocks}, Fig.~\ref{fig:dusty_shocks}). These are visible in post-shock regions as darker regions that run parallel to shock faces in Fig.~\ref{fig:otslice}. 

To see these oscillations more clearly, we take a slice of the simulation at $x = 0.5$ for the snapshot at $t=0.25$. As our simulation has an even-numbered linear resolution, for gas velocity and density, within a slice in $z$ (the direction with translation symmetry) we take the average of the two cells on either side of $x=0.5$. For dust, we consider only those dust grains that reside in the cells on either side of the interface within the slice. To overcome particle noise, we smooth the dust density and velocity profiles using the same procedure we did in Section~\ref{sec:dusty_shocks}. The results of this are shown in Figure~\ref{fig:otslice} for different resolutions at $\mu=0.1$ for $y = 0$ to $0.3$. The analogy to the shock simulation shown in the middle column of Figure~\ref{fig:dusty_shocks} is clear: the morphology of the two simulations is nearly identical. Both simulations exhibit `dust-cyclotron' waves in the post-shock region similar to ion-cyclotron waves, but with dust playing the role of ions and the plasma as a whole taking the part of electrons.

\begin{figure*}
    \centering
    \includegraphics[width=\textwidth]{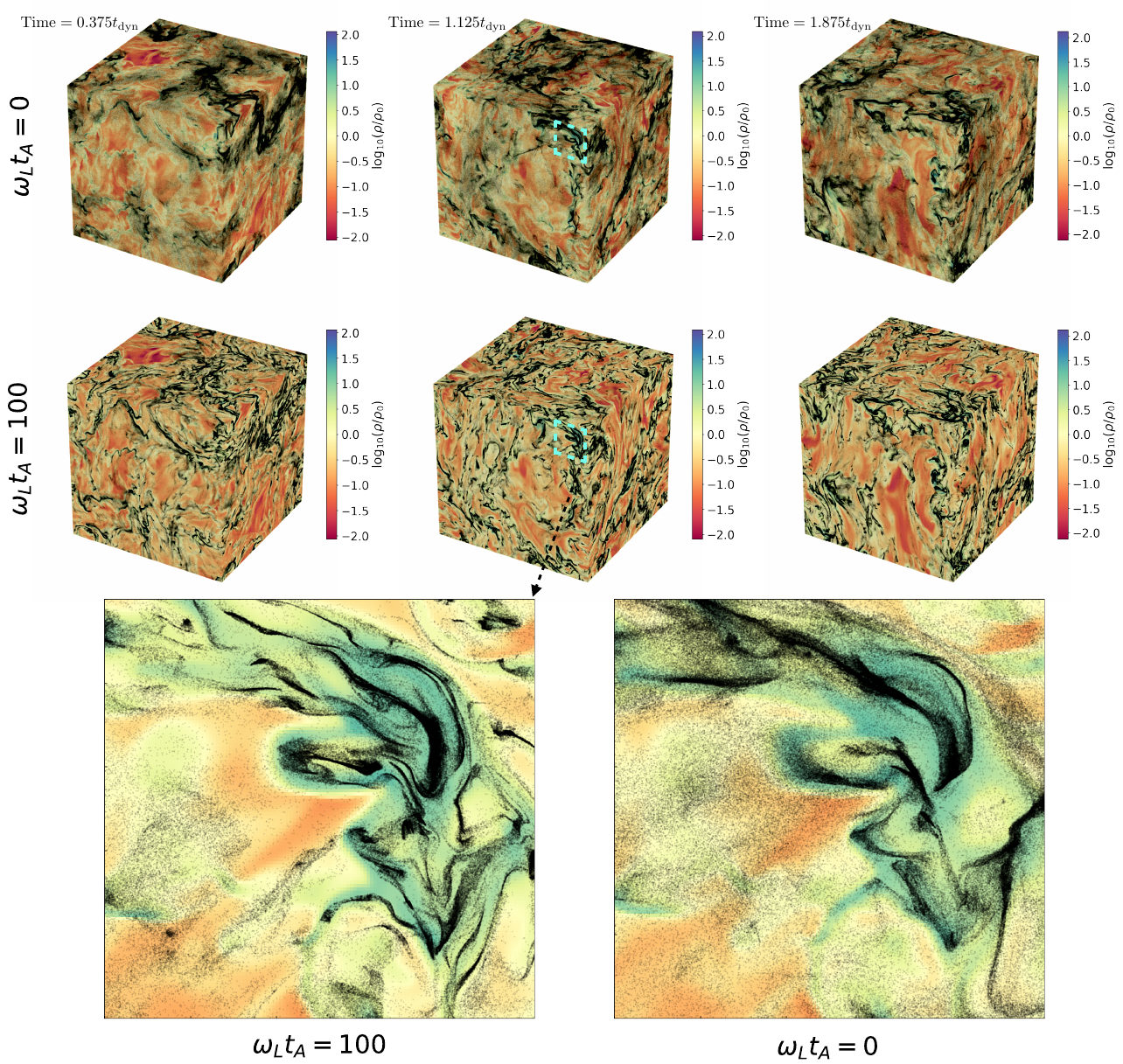}
    \caption{{\bf Top row:} $N_{\rm dust}=N_{\rm gas}=512^3$ resolution simulation of decaying turbulence for uncharged grains ($\omega_{\rm L}t_{\rm A}=0$) at times $t=0.375 t_{\rm dyn}, 1.125 t_{\rm dyn}, 1.875 t_{\rm dyn}$, where $t_{\rm dyn}$ is given by the the box length divided by the standard deviation of the initial velocity. Here, the dust-to-gas mass ratio is 0.01, the initial Mach number is 15, and the initial plasma beta is 0.1, with the magnetic field pointing towards the top of the page. These simulations Epstein drag law (Eq.~\ref{eq:epstein}), with grain-size-parameter $\varepsilon = 0.1$ (cf. secs.~\ref{sec:equations},~\ref{sec:drag}). The face of each cube corresponds to a thin slice of the simulation colored by (log) density with dark points representing the positions of dust grains within that slice. {\bf Second row:} Decaying turbulence for charged grains ($\omega_{\rm L}t_{\rm A} = 100$) at the same times as the uncharged grain simulation. The initial conditions are identical to the uncharged grain simulation. {\bf Bottom row:} A zoom in on slices from the upper right corner highlighted with the dashed cyan box in the middle panels in the top two rows. The image on the left of the last row shows charged grains, while the right shows neutral grains. These plots show the gas density with dust superparticle positions overlain.
    }
    \label{fig:decay_cubes}
\end{figure*}
\subsection{Magnetized Resonant Drag Instability}\label{sec:rdi}

In this section, we present the results of a simulation of the magnetized resonant drag instability (RDI) designed to match (as nearly as possible) the simulation analyzed in SHS19. This is a particularly good test of our back-reaction implementation, because the instability can only grow for $\mu > 0$, and it will only look as it does in SHS19 if both the drag and the Lorentz forces behave properly.

We briefly describe the class of instabilities known as RDIs. \citet{squire2018letter} showed that dust-gas mixtures are generically unstable whenever there is a sustained net drift velocity between the two components. In general, there is at least one instability for every wave mode that the gas supports, e.g. the acoustic RDI for sound waves, the Alfv{\'e}n RDI for Alfv{\'e}n waves, etc.\citep{hopkins2018ubiquitous}

\citet{hopkins2018resonant} performed a linear stability analysis of the acoustic RDI modes. \citet{hopkins2018ubiquitous} then examined RDIs involving charged dust grains, and thus allowing for Lorentz forces to come into play. A follow up to \citet{hopkins2018resonant} was done by \citet{moseley2019non}, where an analysis of the non-linear outcome of the acoustic RDI was studied. One particular set of parameters was chosen for a follow-up study to \citet{hopkins2018ubiquitous}, presented by SHS19. This study was the first to delineate the non-linear outcome of the magnetized RDI. 

We present the results of simulations of the magnetized RDI with initial conditions chosen so that we reproduce the simulation in SHS19. To briefly reiterate their (and our) setup, $\bar{a} = 5$, $\varepsilon = 5$, $\xi = 10$, $\mu = 0.01$, $\beta = 2$, $\gamma = 1$, and $\theta_{\bm Ba} = 87^\circ$, where $\beta$ is the plasma $\beta$, $\gamma$ is the adiabatic index, and $\theta_{\bm Ba}$ is the angle between the magnetic field and the (constant) acceleration vector acting on the dust.\footnote{See Figure 1 in SHS19 for a depiction of this geometry.} $\bar{a}$ is a dimensionless acceleration given by,
\begin{align}
    \bar{a} &\equiv \frac{a \ell_0}{c_{\rm s}^2},
\end{align}
where $a$ is a constant acceleration. Any other parameters here are described in Section~\ref{sec:equations}. 

There are a few small differences in our simulations (aside from the very different numerical methods) compared to SHS19. One is that we add a gravitational force acting on the dust and the gas to balance the force on the dust, so that the gravitational acceleration ${\bm g}$ is given by
\begin{align}
    {\bm g} &= -\frac{\mu}{1+\mu}\bar{\bm a}.
\end{align}
This should not affect the growth or non-linear outcome of the instability, but serves to keep velocities from diverging to very high values that would cause an unnecessarily restrictive time-step. We also use a lower resolution than SHS19, $128^3$ dust particles and gas cells rather than their $256^3$ dust and gas particles. One final difference is that we seed our simulations with white noise in the dust and gas velocities with an amplitude of $10^{-7}$. This contrasts with SHS19 allowing the instability to grow from particle noise. We choose $10^{-7}$ as the amplitude of the initial noise because this is the lowest amplitude that we find our MHD variables are sensitive to, while the particle variables are still subject to noise that makes it hard to pick out a signal below about $10^{-5}$ in velocity and $10^{-4}$ in density. This is one of the advantages of our MHD-PIC method compared to a purely Lagrangian method: we are able to pick up smaller signals in the gas variables and the magnetic field at a similar computational cost. A grid-based code also enables us to employ the constrained transport algorithm for the magnetic field, guaranteeing $\nabla\cdot\bB =0$ is maintained to machine precision at all times.

The growth of the instability versus time can be seen in Figure~\ref{fig:rdi_growth} (cf. fig. 5 in SHS19). Each of the four panels shows the growth of the standard deviations of different variables: the log of the gas and dust densities\footnote{Dust densities are computed by depositing mass onto the grid with the CIC interpolation kernel.}, the three components of dust velocity, gas velocity, and the magnetic field. 

In Figure~\ref{fig:rdi_growth} we also highlight the difference between using TSC and CIC interpolation kernels. The difference is extremely minimal, with TSC growing some constant factor faster than CIC for most variables. For both kernels, we see very clean exponential growth from the low initial seed amplitude up to the slower non-linear growth stage. The slope of these exponential curves closely matches that seen in SHS19, as well as the non-linear growth phase. It is also worth noting that the saturation amplitudes of all 11 variables shown in Figure~\ref{fig:rdi_growth} closely match the amplitudes seen in SHS19. The variables that exhibit less clean growth are the dust density and $x$ and $y$ velocities, due to noise introduced through particle motions. 

Figure~\ref{fig:rdi_cubes} is designed to be as directly comparable as possible to Figure 1 from SHS19. Here, we depict slices along three of the six faces of our TSC simulation at three different times. These times correspond to the linear growth phase, the non-linear growth phase, and the saturated state. The colors on the faces of the cube correspond to the value of the deviation of the magnetic field strength from the mean value (in code units), while dark points correspond to the positions of dust super-particles within a thin slice. The similarity to Figure 1 in SHS19 is striking, with the exact same set of resonant angles appearing, and the saturated state bearing (by eye) many of the same structures. 

\begin{figure}
    \centering
    \includegraphics[width=\columnwidth]{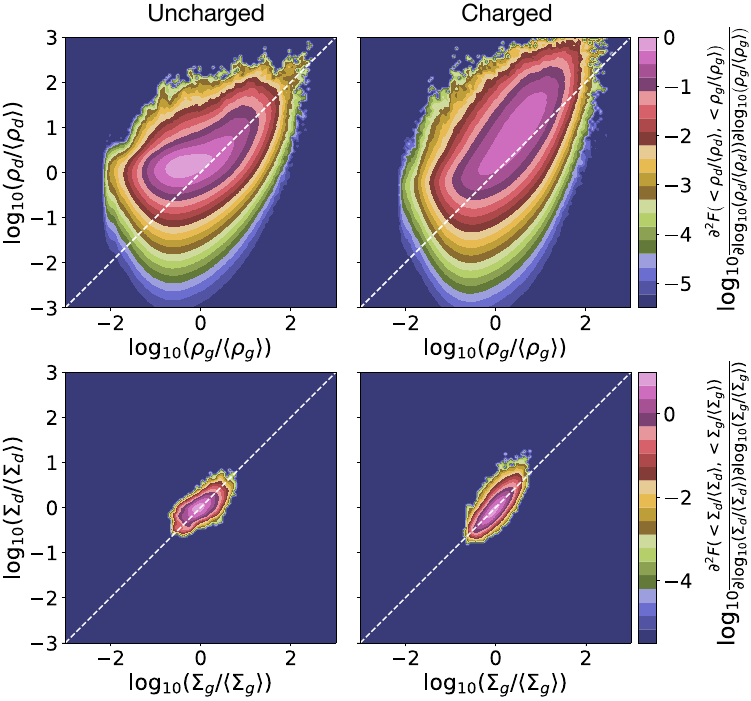}
    \caption{Dust density weighted joint probability density functions of dust density vs gas density (top row) and dust surface density vs gas surface density (bottom row) for uncharged grains (left column) and charged grains (right column) for the simulations shown in Figure~\ref{fig:decay_cubes}, described in detail in Section~\ref{sec:decayturb}. White dashed lines show direct proportionality.}
    \label{fig:density_2dpdfs}
\end{figure}
\begin{figure}
    \centering
    \includegraphics[width=\columnwidth]{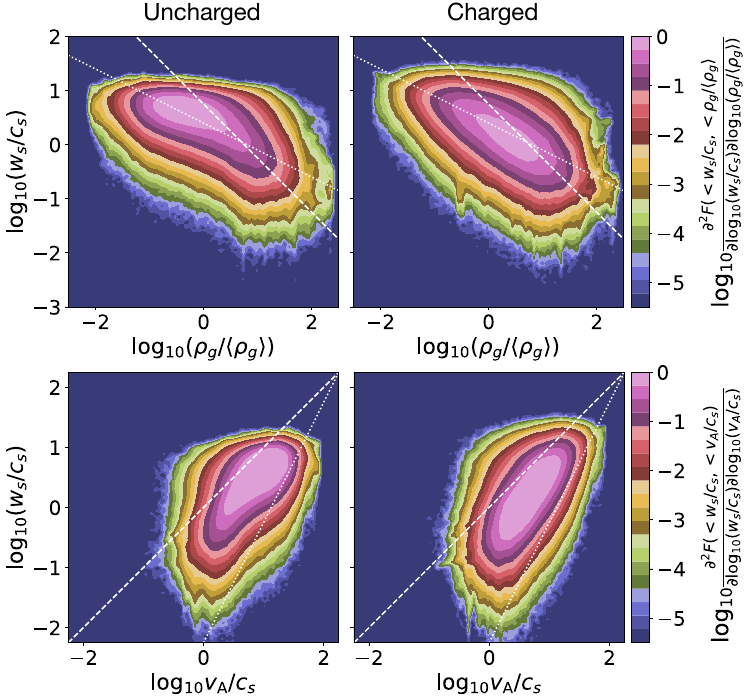}
    \caption{{\bf Top row:} Dust density weighted joint probability density functions of dust drift velocity (in units of the sound speed) vs the logarithm of normalized gas density in our decaying turbulence simulations described in Section~\ref{sec:decayturb}. On the left is our simulation with uncharged grains, on right we use charged grains. The white dashed line shows inverse proportionality between drift velocity $\driftvelmag$ and gas density $\rho_{\rm g}$, while the white dotted line shows $\driftvelmag \propto 1/\sqrt{\rho_{\rm g}}$. {\bf Bottom row:} Dust density weighted joint PDFs of dust drift velocity (in units of the sound speed) vs. the Alfv{\'e}n velocity (in units of the sound speed) for the same simulations. The white dashed lines indicate direct proportionality, while the white dotted line indicates $\driftvelmag \propto v_A^2$.  }
    \label{fig:drift_2dpdfs}
\end{figure}

\begin{figure}
    \centering
    \includegraphics[width=\columnwidth]{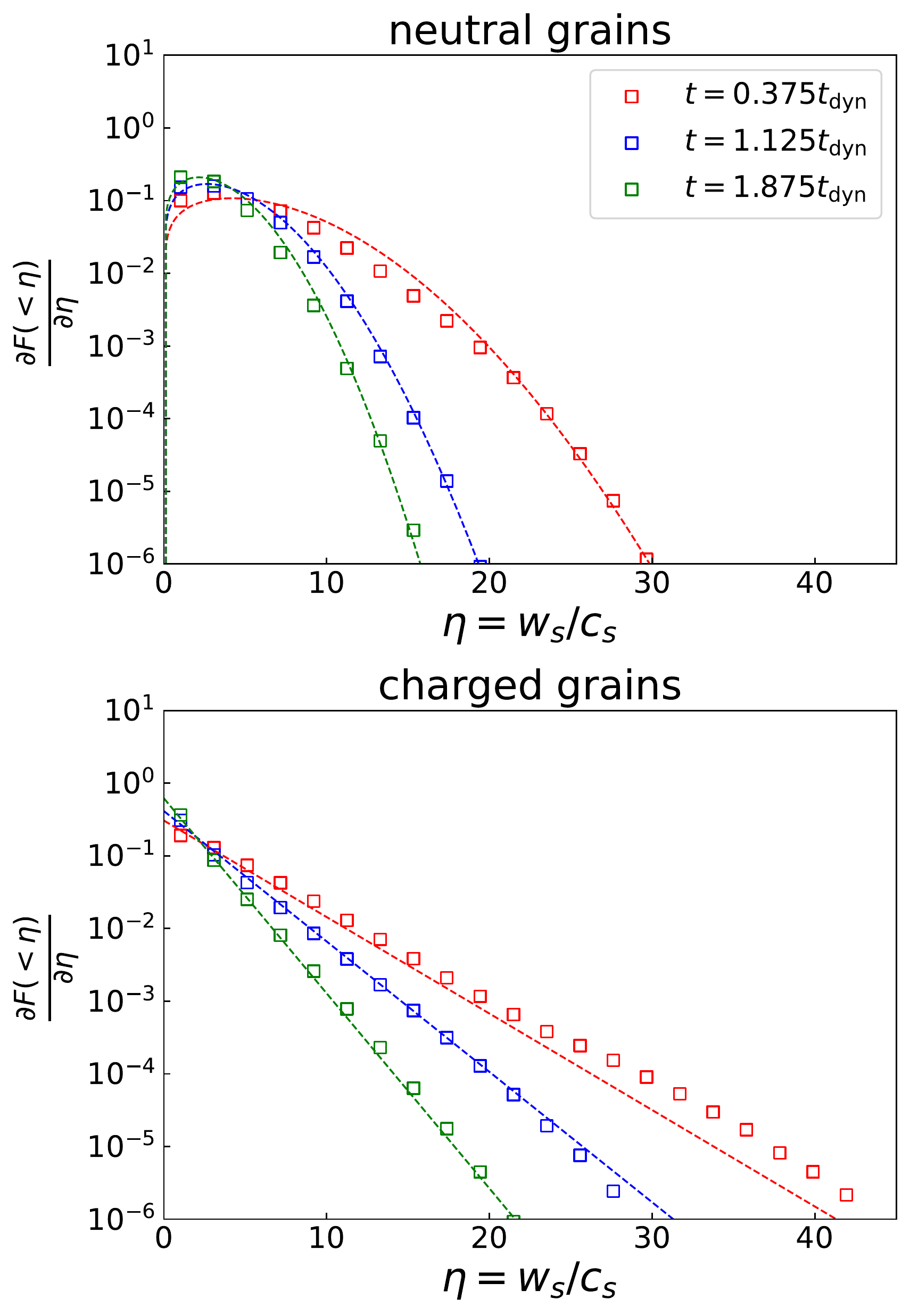}
    \caption{Dust density weighted drift velocity PDFs for charged (bottom) and uncharged (top) grains in the decaying turbulence simulations shown in Figure~\ref{fig:decay_cubes} at times $t=0.375, 1.125, 1.875$ in units of $ t_{\rm dyn}$ (the box length divided by the initial velocity dispersion). Open squares represent the binned simulation data, while dashed lines represent a model that was simply fit by eye to the data at t=1.125. It was not tuned to fit the data at other times, yet it seems to agree with them reasonably well, suggesting that the parameters that are dictating the underlying width are no more complex than the velocity dispersion of the gas and the value of the Alfv{\'e}n velocity. These simulations use the Epstein-Baines drag law (Eq.~\ref{eq:kwok}). Notice the exponential tail in the PDF for charged grain drift velocities, while the uncharged grains follow more nearly a (modified) gaussian distribution. While the typical drift velocity is larger for uncharged grains, the charged grains are capable of reaching larger drift velocities.}
    \label{fig:drift_pdfs}
\end{figure}

\subsection{Decaying turbulence}\label{sec:decayturb}
In this subsection, we present the results of isothermal decaying turbulence simulations meant to roughly correspond to 1-2 micron dust grains in a 20 pc patch of cold neutral medium with a density $n_{\rm H} = 30 \hspace{0.1cm}{\rm cm}^{-3}$. The size these grains correspond to exactly depends upon the composition of the grains. One simulation is run including Lorentz forces on the grains, the other without, i.e. for one simulation, the charge to mass ratio is $\xi = 0$, and the other has $\xi = 100$. $\xi=100$ corresponds roughly to a charge on the grains of $100 \lesssim Z_{\rm d} \lesssim 250$ depending upon the grain composition. This is compatible with charging due to photoelectric emission in a typical patch of cold neutral medium (CNM) \citep{Yan+Lazarian+Draine_2004}.

The dust grains in both those simulations have their grain size parameter $\varepsilon = 0.1$. The dust to gas mass ratio in all of these simulations is chosen to be $\mu = 0.01$, the initial plasma $\beta = 0.1$, the initial 3D sonic Mach number is $\mathcal{M} = 15$. For $T=100\K$ and a mean molecular weight $\mu_{\rm mol} = 1.49$, the magnetic field strength is $B = 9.8\muG$ and the initial 3D velocity dispersion is $11.1 {\rm km/s}$. Though this velocity dispersion seems somewhat high, it is important to keep in mind that the turbulence rapidly decays to a more reasonable dispersion in line with the standard size-linewidth relation \citep{Larson_1981, Solomon+Rivolo+Barrett+Yahil_1987}.

Maps of the gas density with dust overlain are shown in Figure~\ref{fig:decay_cubes}, along with plots where we zoom in on a small slice of the simulation in one corner of each. All of these simulations are run with the exact same initial conditions at a resolution of $N_{\rm dust} = N_{\rm gas} = 512^3$ dust super-particles and gas cells. We initialize the simulations with large-scale modes between wavenumbers one and two with the desired sonic Mach number and a uniform gas density. Dust is initially chosen to be moving perfectly with the gas, with one dust super-particle initialized at the center of each cell.

We have also run simulations with and without the supersonic Baines correction to explore the impact of a variable drag time on the drift velocity PDFs (Fig.~\ref{fig:drift_pdfs}). Our results do not qualitatively change, though there is a pronounced effect from the stronger coupling imparted by the Baines correction.

\subsubsection{Dust spatial distribution and dust-to-gas mass ratio}\label{sec:dust-to-gas}
In Figure~\ref{fig:density_2dpdfs}, we present joint PDFs of the dust and gas densities, as well as the dust and gas surface densities for our decaying turbulence simulations with the Baines correction to the drag law. Uncharged grains, while reasonably well-coupled to the gas in the high-density tail, are relatively uncoupled to gas at or below the mean density. Charged dust grains, on the other hand, are better coupled to the gas in the low-density limit. This is presumably because the coupling is not being provided by the drag, but instead through the magnetic field. On the other end of the distribution, in the high-density limit, the charged grain density is actually super-linear with gas density, while it is merely linear for charged grains. When averaging over a column to get the joint surface density PDFs, we see that most of the scatter in the densities is averaged out. The tendency of charged dust to couple more tightly to gas as compared to uncharged dust is evident here as well.

It is also the case that charged dust grains reach higher dust densities than uncharged grains in these simulations. We suspect this is directly related to dust density oscillations in post-shock regions that we have seen in our dusty Orszag-Tang vortex simulations (cf. figs.~\ref{fig:dustyot},~\ref{fig:otslice}, Section~\ref{sec:orszag-tang}) as well as our shock simulations (cf. Fig.~\ref{fig:dusty_shocks}, Section~\ref{sec:dusty_shocks}). In fact, looking at the spatial distribution of dust grains in Figure~\ref{fig:decay_cubes} and comparing the charged-grain to the uncharged-grain simulation, it is clear that grains are more strongly clustered when they are charged. Looking at the zoomed-in region in the bottom two rows, it is also clear that there is more small-scale structure in the dust distribution for charged grains as compared to uncharged grains. It appears that for many lone, broad features in uncharged dust, we see corresponding dust sheets in charged dust. Where a shock might produce a single broad smear in uncharged dust, it produces many gyro-scale density oscillations in charged dust. 

\subsubsection{Dust velocity PDFs and shattering rates} \label{sec:shattering}
\begin{table*}
\centering
\begin{tabular}{||c| c c c | c c c||} 
 \hline
variable & &charged& & &neutral& \\ [0.5ex]
 Time $t/t_{\rm dyn}$ elapsed &$0.375$ &$1.125 $&$1.875 $ & $0.375$&$1.125$&$1.875 $\\
 \hline\hline
 $B_{\rm rms}$($\mu$G)  &17.5 & 14.5& 12.2 &17.4 & 14.5&12.1\\
$v_{{\rm A},{\rm rms}}$(km/s) & 8.31 & 9.65& 7.98 &8.31  &9.62 & 7.97\\
$u_{\rm rms}$(km/s)& 6.56 & 4.18 &3.38 & 6.55 & 4.17 & 3.37\\
$v_{\rm rms}$(km/s) & 7.33 & 4.67 & 3.68 & 6.81 & 3.81 & 2.97 \\
$w_{{\rm s},{\rm rms}}$(km/s) & 3.85 &2.39 & 1.67 & 4.54 &3.16& 2.38 \\
 \hline\hline
 $a_{\rm gr}$ - graphite (\um) & 1.96 & - &- & -&- &-  \\
 $a_{\rm gr}$ - silicate (\um) & 1.31 & - &- &- & -&- \\
 $\tau_{\rm gr}$ - graphite (Myr) & 3.09 & 14.06 & 47.94  & 2.50 & 11.95 & 27.33 \\
 $\tau_{\rm gr}$ - silicate (Myr) & 16.75  & 83.48 & 251.5 & 13.24 & 70.12 & 138.3 \\
$\gamma_{\rm rel}$ - graphite  & 0.81& 0.85 & 0.57&- & -&-  \\
$\gamma_{\rm rel}$ - silicate  & 0.79 & 0.84 & 0.55& -& -&-  \\ [1ex] 
 \hline
\end{tabular}
\caption{{\bf Top:} Various statistics for our $512^3$ resolution decaying turbulence simulations with Epstein-Baines dust drag for charged (left) and neutral (right) grains. {\bf Bottom:} {Grain sizes that we use in the simulation, assuming either a pure graphite or silicate composition as well as the mean grain lifetime $\tau_{\rm gr}$ (cf. Eq.~\ref{eq:tau}) and the factor by which the shattering rates are altered by grain charge $\gamma_{\rm rel}$ (depending upon the composition of grains) (cf. Eq.~\ref{eq:rel}). These quantities are discussed in Section~\ref{sec:shattering}.} We compute these quantities assuming all collisions are with grains that are perfectly coupled to the gas. Note that whatever composition is assumed for grains, charge offers the grains relative protection from shattering through the enhanced mean coupling to the gas, with the shattering rate being \textit{reduced} by 15–45\% depending on grain composition and the simulation snapshot.}
\label{tab:shat}
\end{table*}

Figs.~\ref{fig:drift_2dpdfs} and \ref{fig:drift_pdfs} show the probability distributions of dust drift velocities in our simulations. Figure~\ref{fig:drift_2dpdfs} shows the joint-PDF of drift velocity and gas density (top row) and drift velocity and Alfv{\'e}n speed (bottom row) across our two simulations. Figure~\ref{fig:drift_pdfs} shows the PDFs of drift velocity for two different simulations using  Epstein-Baines drag (Eq.~\ref{eq:kwok}), with and without the Lorentz force acting on dust grains (bottom and top panels, respectively in the figure). 

Uncharged dust grains seem to follow quasi-gaussian profiles, while charged grains seem to follow almost perfect exponential profiles. The curves in the top panel (uncharged grains) of Figure~\ref{fig:drift_pdfs} correspond to grain drift-speed distributions where,
\begin{align}
    p(w_{\rm s}) &\propto \sqrt{w_{\rm s}}\exp\left(-\frac{w_{\rm s}^2/c_{\rm s}^2}{2\times0.44\mathcal{M}_{\rm rms}^2}\right).\label{eq:supersonic_nocharge}
\end{align}
Here, $\mathcal{M}_{\rm rms}\equiv u_{\rm rms}/c_{\rm s}$ is the root-mean-square sonic Mach number. The curves in the bottom panel (charged grains) correspond to,
\begin{align}
    p(w_{\rm s}) &\propto \exp\left(\frac{-w_{\rm s}/v_{\rm A,rms}}{0.033\mathcal{M}_{\rm rms}}\right).\label{eq:supersonic_charge}
\end{align}
Here, $v_{\rm A,rms}$ is the root-mean-square Alfv{\'e}n velocity. These curves are simply by-eye approximations to the profiles at time $t=1.125t_{\rm dyn}$, not linear regressions. After the curves were fit by eye to the snapshot at $t=1.125t_{\rm dyn}$, we then adjusted the profiles according to physically motivated guesses for how we thought the profile should scale with $\mathcal{M}_{\rm rms}$ and $v_{\rm A, rms}$. 

In trying to explain the shapes of these distributions, we can turn to the Fokker-Planck equation. For uncharged dust grains, the diffusion comes purely from the change in the underlying gas velocity as the grain moves through it. In that scenario, the equilibrium distribution of dust grain velocities will mirror that of the gas, and the distribution of grain drift velocities will have a similar shape, but have a width that depends upon the strength of the drag. 

For charged grains, the distribution is harder to understand, as cross-correlations between the Lorentz force and the gas velocity are important. However, as the magnitude of the Lorentz force depends upon the grain drift velocity, it is clear that grains with higher drift velocities will experience a higher level of diffusion. This increased diffusion means that fast grains are more likely to wind up diffusing to even higher velocities than slow grains. Qualitatively this explains the exponential shape of the distribution. As the grains in this simulation have $\omega_{\rm L}t_{\rm s} \gg 1$, Lorentz forces should dominate over drag forces most of the time, and so the Alfv{\'e}n velocity together with the gas velocity dispersion (sonic Mach number) should determine the width of the profile. In contrast, the drift velocity PDF of neutral grains should not depend strongly on the Alfv{\'e}n velocity because they have no direct coupling to the magnetic field. 

We now consider the rate of erosion $\Gamma_{\rm shat}$ of our simulated (``target'') grain population due to collisions with smaller (``impactor'') grains assumed to be perfectly coupled to the underlying fluid, as well as the ratio of these rates between the two (otherwise identical) simulations, $\gamma_{\rm rel}$, and the typical target grain lifetime, $\tau_{\rm gr}$. Values for $\gamma_{\rm rel}$ and $\tau_{\rm gr}$ are given in Table~\ref{tab:shat}. We now explain how these values are estimated. 

We express the overall average shattering rate as

\begin{equation}
    \Gamma_{\rm shat} = \frac{\pi a_{\rm tar}^2\mu}{v_{\rm shat}^2} \left\langle \rho w_{\rm s}(>v_{\rm shat})^3\right\rangle\label{eq:shat},
\end{equation}
where $a_{\rm tar}$ is the radius of the ``target'' grains (the ones we simulate directly), $v_{\rm shat}$ is the collision velocity above which grains are taken to erode, and the ensemble average on the right is taken over dust superparticle positions. The values of $v_{\rm shat}$ and $\rho_{\rm d}^{\rm i}$ that we adopt are taken from \citet{Jones+Tielens+Hollenbach_1996}. These values are 1.2 and 2.7$\,$\kms$\,$ for $v_{\rm shat}$ and 2.2 and 3.3$\,{\rm g}/{\rm cm}^3$ for graphite and silicate, respectively. 

To compute the target grain radius using our grain size parameter $\varepsilon = 0.1$, we have
\begin{align}
    a_{\rm tar} = 0.1 \ell_0\rho_0/\rho_{\rm d}^{\rm i}.
\end{align}
We assume that $\rho_0 = 7\times10^{-23}\,{\rm g/cm}^3$ is the mean gas density in our simulation, and $\ell_0 = 20\,{\rm pc}$ is the length of one side of our simulation volume. In order to scale grain velocities to physical units, we assume a gas temperature of 100\K, corresponding to a sound speed of $c_{\rm s} = 0.74\,{\rm km/s}$ given a mean molecular weight of 1.5. 
Then, dividing $\Gamma_{\rm shat}$ by the target grain's mass, we can compute the inverse of the grain lifetime, $\tau_{\rm gr}$ as
\begin{equation}
    \tau_{\rm gr}^{-1} \equiv \frac{3\mu}{4 a_{\rm tar} \rho_{\rm d}^{\rm i} v_{\rm shat}^2} \left\langle \rho w_{\rm s}(>v_{\rm shat})^3\right\rangle. \label{eq:tau}
\end{equation}
The relative shattering rate is defined by,
\begin{align}
    \gamma_{\rm rel} \equiv \frac{\Gamma_{{\rm shat,}\,{\rm charged}}}{\Gamma_{{\rm shat,}\,{\rm neutral}}} = \frac{\tau_{{\rm gr},\,{\rm neutral }}}{\tau_{{\rm gr},\,{\rm charged}}}.\label{eq:rel}
\end{align} 
We arrive at these expressions as follows. First, assume spherical grains. Next, assume a composition for grains, either graphite or silicate. The composition will determine both the material density of dust grains, $\rho_{\rm d}^{\rm i}$ (which will determine how large a grain is in physical units), as well as the collision velocity above which grains erode, $v_{\rm shat}$. The amount of material removed in a collision with relative velocity $w_{\rm s}\ge v_{\rm shat}$ is taken to be proportional to the energy of the impactor, while the rate at which these impacts occur is proportional to the drift velocity as well as the density of impactors (assumed to be proportional to $\rho$). The overall shattering rate is then $\Gamma_{\rm shat} \propto \rho w_{\rm s}^3$, provided that $w_{\rm s} > v_{\rm shat}$. 

The values for grain lifetimes $\tau_{\rm gr}$ and the relative grain shattering rate $\gamma_{\rm rel}$ between simulations using charged and neutral grains shown in Table~\ref{tab:shat} show that charge appears to offer these large grains relative protection from erosion and shattering. This is because the r.m.s. drift velocity of charged grains is lower (Tab.~\ref{tab:shat}), though the drift velocity distribution of charged grains extends to higher values (cf. Fig.~\ref{fig:drift_pdfs}). Interestingly, the actual r.m.s. velocities (relative to the ``lab'' frame) are higher for charged grains than for uncharged grains. Thus, it must be that charged dust is somehow concentrating itself into significantly higher fluid velocity regions than uncharged dust.


The results from this section are not conclusive, and more detailed analysis is needed. As we have used decaying turbulence rather than driven, the system is out of statistical equilibrium. The impact of this on our results is unclear. As well, as we only use a single set of initial conditions and a single grain size, the impacts of the turbulent sonic and Alfv{\'e}n Mach numbers, plasma beta, and grain size and charge are also unclear, although examining the profiles at different times has given us some indication as to how the grain drift velocity PDFs should depend upon some of these parameters.

\section{Discussion}\label{sec:discussion}
\subsection{Grain acceleration}
The primary focus of this paper has been to present and validate our implementation of MHD-PIC dust dynamics in RAMSES. Even so, the numerical tests we have chosen have been selected in part because they are physically interesting and help us to better understand the process of dust acceleration in the interstellar medium. 

The two main mechanisms we have examined in this paper are those of gyro-resonance and shock acceleration. We have not investigated other acceleration mechanisms in a simplistic setup (e.g. \S~\ref{sec:dusty_shocks},~\ref{sec:dusty_alfven}), though other mechanisms may well be important and warrant investigation in future work. Of course, more complex and non-linear mechanisms are present implicitly in our decaying turbulence simulations, though they are hard to isolate from one another or identify. Both gyro-resonance and shock acceleration have been investigated in detail in past work \citep{Lazarian+Yan_2002, Yan+Lazarian_2003, Yan+Lazarian+Draine_2004, Yan_2009, guillet2009shocks, Hoang+Lazarian+Schlickeiser_2012}.

\citet{Yan+Lazarian+Draine_2004} examined the phenomenon of grain acceleration by gyroresonance with magnetosonic waves using quasi-linear theory (QLT). They concluded that gyroresonance with fast waves dominated grain acceleration, accelerating grains bigger than $0.1\,$\um$\,$ to velocities $\gtrsim 1\,$ \kms$\,$ in cold neutral medium and $\gtrsim 20\,$ \kms$\,$ in warm neutral medium. The velocities they predicted in cold neutral medium are close to the grain shattering/eroding velocities cited in \citet{Jones+Tielens+Hollenbach_1996}, while those in warm neutral medium are fast enough to vaporize grains.

There are several caveats to their results that beg further investigation, both related to fundamental assumptions of their theory. The basic assumptions of the theory are that (i) the guiding center of grains moves in a regular trajectory along a uniform magnetic field, and (ii) the underlying turbulence is sub-Alfv{\'e}nic. Assumption (ii) may often not be the case in molecular clouds \citep[e.g.][]{padoan2004average}. If assumption (ii) is false, then assumption (i) will also often be untrue, for the magnetic field may vary on a length-scale smaller than the grain gyro-radius, especially for large grains. Another caveat is that they predict specific values for grain velocities. These values are ensemble averages, and do not give information about the underlying velocity distribution. The shape of the underlying distribution can impact the shattering rate, as we have seen in Table~\ref{tab:shat}, and so this is not necessarily negligible information. Given that our simulations are super-Alfv{\'e}nic, we do not expect the results from \citet{Yan+Lazarian+Draine_2004} to apply straightforwardly to our simulations in Section~\ref{sec:decayturb}, though the mechanisms that they explore may indeed be present in some regions of the turbulent flow. As well, the grains we investigate are quite large, and we do not probe deep into the sub-Alfv{\'e}nic regime where their theory would apply directly.

\citet{Yan_2009} later revisited the problem of grain acceleration, building upon \citet{Yan+Lazarian+Draine_2004} and accounting for betatron acceleration, where compressible MHD modes produce a non-zero $\nabla\times\bE$ aligned with the magnetic field, accelerating grains in the direction across the field. \citet{Yan_2009} concluded that this mechanism is dominant when grains move super-Alfv{\'e}nically (as in our simulations in Section~\ref{sec:decayturb}), and subdominant to gyroresonance when grains move sub-Alfv{\'e}nically. While the same limitations of QLT still largely apply, the betatron mechanism could be part of what is accelerating grains in the simulations of Section~\ref{sec:decayturb}. However, this mechanism only works under circumstances where the first adiabatic invariant (the magnetic moment $\mu_B = m v_{\bot}^2/(2B)$) is conserved. That is, when grains undergo stable gyro-orbits, or when the gyro radius $r_{\rm L}$ is much less than the scale $\ell_B$ over which the magnetic field varies. While this may be true in some regions of the simulations presented in Section~\ref{sec:decayturb}, these assumptions will often be violated, e.g. in shocks, or when the grains have acquired a large velocity already so that $r_{\rm L}$ is appreciable.

\cite{Hoang+Lazarian+Schlickeiser_2012} updated the theory of \citet{Yan+Lazarian+Draine_2004} with a new non-linear theory (NLT) for gyroresonance and adding an additional acceleration mechanism termed transit-time damping (TTD). NLT allows for fluctuations of the grain guiding center by assuming that they are gaussian distributed. This effectively broadens the resonance condition for gyroresonance. TTD amounts to grains ``seeing'' more mirror-acceleration from compressive modes in the direction of wave propagation than in the opposite direction. The authors found that TTD can be dominant to gyroresonance by an order of magnitude for grains larger than 0.05$\,$\um$\,$, while also being more efficient than the betatron acceleration studied by \citet{Yan_2009}, so long as grains move super-Alfv{\'e}nically. TTD may indeed be the mechanism by which the charged grain simulation of Section~\ref{sec:decayturb} are accelerated, but further study is needed to confirm this.

Regarding the topic of shock acceleration, Guillet, Pineau des For{\^e}ts, Jones, Anderl, and Flower (variously) wrote a series of papers \citep{guillet2007shocks,guillet2009shocks,guillet2011shocks,anderl2013shocks} exploring the topic. As discussed in Section~\ref{sec:dusty_shocks}, of these papers, the one whose results are most immediately comparable to our simulations in Section~\ref{sec:dusty_shocks} is \citet{guillet2009shocks}, wherein the authors investigated dust destruction in J-shocks. We do not investigate the destruction of dust in shocks directly, but rather, grain acceleration. To reiterate the results of Section~\ref{sec:dusty_shocks}, we find that for grains initially at rest, their drift velocity relative to the gas can never be larger than it is at the shock front if the grain only crosses the shock front once. Further, provided the grains only cross the shock front once, Lorentz forces do not serve to further increase the velocity of grains in shocks, merely to cause gyro-motion. However, the \textit{relative} velocity of similarly sized grains may be enhanced via this mechanism, allowing for collisions of grains through gyromotion that otherwise would occur less often with hydrodrag alone. 

The simplistic setup in Fig.~\ref{fig:dusty_shocks} (cf. Sec.~\ref{sec:dusty_shocks}) also neglects the possibility of first order Fermi acceleration (FOFA) \citep{Fermi_1949}. FOFA requires the presence of magnetic inhomogeneities, with a more magnetically inhomogenous medium being more effective in accelerating particles. As well, when particles are substantially super-thermal, the mechanism is more effective. This last condition is met in our turbulence simulations simply by virtue of the imperfect dust-gas coupling, as we have seen in our decaying turbulence simulations. As well, magnetic inhomogeneities are ubiquitous in MHD turbulence simulations like ours. Thus, FOFA could very well manifest in the simulations of Section~\ref{sec:decayturb}. However, it cannot in the shock simulations of Figure~\ref{fig:dusty_shocks} (also discussed in Sec.~\ref{sec:dusty_shocks}) without either a higher Alfv{\'e}n Mach number, or more complex initial conditions and upwind/downwind magnetic field geometry. We demonstrate mild FOFA in Fig.~\ref{fig:mach6} by increasing the Alfv{\'e}n Mach number to $\mathcal{M}_{\rm A} = 6$ instead of $\mathcal{M}_{\rm A} = 3$. 

As stated in Section~\ref{sec:decayturb}, we have chosen to model large, 1-2 micron grains in our simulations. While most of the dust mass in the ISM is thought to be contained in grains that are $\lesssim 0.1\,\mu{\rm m}$, larger grains (even up to $10\,\mu{\rm m}$) have been found trapped in meteorites at the level of a few parts per million \citep[e.g.][]{Anders+Zinner_1993,Takigawa_2018, Heck_2020}. 
Using Ne isotopes produced by galactic cosmic rays, \citet{Heck_2020} found that a majority of large grains ($\gtrsim 1\,\mu{\rm m}$) trapped in meteorites were exposed to the ISM for less than 300 Myr, with a minority (8\%) being exposed for > 1 Gyr, shorter lifetimes than expected. \citet{Takigawa_2018} studied a single 1.4 $\mu{\rm m}$ presolar grain that showed evidence of a single collision throughout its lifetime that formed high-Mg domains, one rough surface, and a cavity. The grain shapes observed by \citet{Heck_2020} and \citet{Takigawa_2018} are remarkably smooth, suggesting little in the way of gas-grain or grain-grain collisions/erosion. 

In the context of our work, we would interpret this as meaning that these large grains have had relatively low average drift velocities throughout their < 1 Gyr lifespans. This is puzzling, as the mean grain lifetime in our simulations differs from these numbers significantly, with the lifetimes we calculate ranging from 2.5 - 250 Myr, depending upon the snapshot, simulation, and grain composition. 

However, it should be noted that we only have considered a single phase of ISM. In lower density ISM phases, grains will experience significantly lower erosion rates, even as the surrounding medium is hotter. As the ISM is well-mixed \citep{Weingartner+Draine_1999}, and as only a small fraction of ISM mass is in cold neutral medium (such as that which we model here), grains will spend much of their time at lower densities and thus experience lower erosion rates.
The existence of presolar grains with long ($\sim$Gyr) exposure times which appear to have escaped significant interstellar erosion seems inconsistent with the mean lifetimes $< 250\,$Myr for the similarly sized grains in our decaying turbulence simulations. Perhaps there is a population of grains whose typical drift velocity remains low indefinitely. In future work, we plan to evaluate erosion rates along grain trajectories, rather than only instantaneous rates averaged over all grains.  It is also the case that the \citet{Heck_2020} and \citet{Takigawa_2018} samples are subject to survivorship bias.

\subsection{Turbulent grain concentration}
We have also touched on the topic of turbulent grain concentration (and by extension, spatial grain size-sorting) in Section~\ref{sec:dust-to-gas}. \citet{hopkins2016fundamentally} and \citet{lee2017dynamics}  have explored this topic with simulations that are very comparable to ours, with the former studying neutral grains and the latter studying charged grains. They used driven MHD turbulence simulations with dust treated as massless particles whose dynamics were followed in real-time. The authors neglected dust back-reaction (unimportant except for the limited number of regions where they found the dust to gas mass ratio $\mu > 1$), though their code \citep[GIZMO,][]{Hopkins_2015} now has a self-consistent implementation of dust back-reaction \citep[e.g.][]{moseley2019non,seligman2019non, steinwandel2021optical}. Like our simulations, the authors allow for Epstein-Baines drag (Eq.~\ref{eq:kwok}), and in \citet{lee2017dynamics} specifically, Lorentz forces on grains. 

\citet{hopkins2016fundamentally} identified three limiting cases for dust concentration. In the first, the grain size parameter $\varepsilon \gtrsim 1$, and dust is effectively decoupled from gas motions, remaining more or less near the mean density at all times. In the second, $1/\mathcal{M}^2 \lesssim \varepsilon \lesssim 1$ ($\mathcal{M}$ is the turbulent Mach number), and dust grains are concentrated in non-trivial ways in many regions. In the third, $\varepsilon \ll 1/\mathcal{M}^2$, and dust grains are effectively coupled to the gas everywhere, with $\rho_{\rm gas} \propto \rho_{\rm dust}$. The simulations we have presented here are firmly in the second regime, as we have $\mathcal{M} \lesssim 20$ and $\varepsilon = 0.1$ in Section~\ref{sec:decayturb}. \citet{hopkins2016fundamentally} had a simulation where $\mathcal{M} \approx 10$ and $\varepsilon = 0.1$, and found results that are very similar to those we see in our uncharged dust simulations. Like us, they observe uncharged dust with $\varepsilon = 0.1$ as being less coupled in low density regions and coupled in the high density regions. On the other hand, we observe dust density fluctuations down to a lower value and maximum density fluctuations about 1-2 orders of magnitude lower than they observed. Some of this difference is attributable to their post-processing technique for obtaining dust and gas densities at the location of dust particles. 

The simulations presented in \citet{lee2017dynamics} include Lorentz forces on grains. They also find that this effectively couples grains to the gas in lower densities where hydrodrag failed to do so. They do not find the same super-linear relationship between dust density and gas density, though this should not be troubling: our simulations are decaying, and thus not in statistical equilibrium. Should the effect persist for driven turbulence simulations, it is possible that it is due to the nature of treating gas with an Eulerian method and dust with a Lagrangian method. If this is the case, however, we might also expect to observe this same super-linear scaling for neutral grains, though not necessarily. A final possibility is that it is due to differences in our numerical methods for either dust or gas relative to those used in \citet{hopkins2016fundamentally} and \citet{lee2017dynamics}.

We may also compare our results to those of \citet{Beitia+Gomez+Vallejo_2021}, who ran 2D MHD simulations treating dust as massless Lagrangian particles to examine the formation of dusty filaments in molecular clouds. While the fact that their simulations are 2D implies much stronger filamentation than should occur in 3D, the conclusion that (single-sized) charged dust grains should clump more strongly than uncharged dust grains should carry through into 3D. We find this to be true in our decaying turbulence simulations as well, though it remains to be seen if this is true with a range of grain sizes.

The results of \citet{hopkins2016fundamentally}, \citet{lee2017dynamics}, \citet{Beitia+Gomez+Vallejo_2021} and our simulations in Section~\ref{sec:decayturb} apply to grains of specific sizes. This implies that spatial grain size-sorting should occur for grains as a function of which of the three limiting cases they fall into. This could have implications for the local dust to gas mass ratio as well, though to say this conclusively, simulations with a range of grain sizes and ideally many dust particles per gas cell (for more robust statistics in low density regions) would be necessary. 
\section{Summary}\label{sec:summary}
In this paper, we have presented and extensively validated our implementation of MHD-PIC dust dynamics in the fluid dynamics code RAMSES. Our code is currently one of only a few that can simultaneously model dust drag, Lorentz forces, and back-reaction self-consistently. In order to solve the coupled dust-gas system cell-by-cell, we employ the `particle-mesh-back-reaction' method of \citet{yang2016integration}. This method also gives better convergence rates for dust-gas instabilities. 

Dust-gas forces are treated in an operator-split way. The Lorentz operator that we employ is $A-$stable, second order, and symplectic, and so is ideally suited to the Lorentz force. Importantly, unlike the popular Boris algorithm for Lorentz forces on particles \citep{boris1970relativistic}, we do not operator-split the effects of the electric and magnetic fields, but instead consider their joint action upon the dust-gas drift velocity. The two drag operators we employ (first and second order) are $L-$stable, and thus ideally suited to modeling drag in stiff regimes where the stopping time is less than the time-step or where the concentration of dust solids is large. These two properties together mean that where the dust-to-gas  mass ratio $\mu \gg 1$, we do not need to artificially alter either the stopping time $t_{\rm s}$ or the Larmor frequency $\omega_{\rm L}$ to obtain reasonable results in these (statistically rare) regions. Both of our solution operators work on the dust-gas \textit{drift velocity} rather than simply on the dust velocity alone. Doing this gives us the previously mentioned stability properties that would otherwise not be possible for non-zero dust-to-gas mass ratios using more common methods. While other methods can work (e.g. solving for Lorentz and drag updates simultaneously), with the appropriate choice of timestep, for $\mu>0$ other methods will not be simultaneously energy conserving in the limit of zero drag (resulting in decaying orbits), $A-$stable in the limit of zero drag, and $L-$stable when drag is present.

To validate our methods, we have chosen interesting yet simple test problems that also serve to build physical intuition for the problem of dust acceleration. These tests include simulations of damped bulk gyromotions of dust grains, dusty shocks, dusty Alfv{\'e}n waves, the dusty Orszag-Tang vortex, the magnetized resonant drag instability, and 3D decaying MHD turbulence. We have done preliminary analysis of grain acceleration in decaying turbulence, as well as the concentration of solids. Our work is consistent with that of previous authors, such as \citet{guillet2009shocks, hopkins2016fundamentally, lee2017dynamics, seligman2019non}. 
Importantly, our dusty turbulence simulations go beyond dust acceleration models discussed in \citet{Yan+Lazarian+Draine_2004, Yan_2009} and \citet{Hoang+Lazarian+Schlickeiser_2012} by probing a regime that violates fundamental assumptions underlying those theories. {Interestingly, we have found that, at least in this simplified setup, Lorentz forces seem to offer grains relative protection from shattering (as compared to the case where they are absent).} More thorough investigation of the phenomena of dust acceleration and turbulent concentration will be explored in forthcoming work \citep[][in prep.]{MoseleyInPrep}. In addition, we have estimated grain lifetimes in our simulations and find they range from 2.5 - 250 Myr for micron-sized grains. This is significantly shorter than those estimated using cosmic-ray-produced neon isotopes found in presolar grains \citep{Heck_2020}. Furthermore, presolar grains have been found to have smooth, relatively unpitted surfaces, a finding inconsistent with the high erosion rates we estimate \citep{Takigawa_2018, Heck_2020}. 

{To summarize:}
\begin{itemize}
    \item {We have implemented MHD-PIC methods for dust in the astrophysical fluid code RAMSES.}
    \item {Our methods are stable even when the dust-to-gas mass ratio $\mu\gg1$, or $\Delta t\gg t_{\rm s},t_{\rm L}$.}
    \item {We have validated our code using a number of physically interesting test problems that demonstrate various grain acceleration mechanisms, as well as test our implementation of grain back-reaction.}
    \item {Our final test, decaying 3D MHD turbulence including dust of a single grain size, shows that grains can be dynamically concentrated to high density as well as accelerated to high velocities by the underlying turbulence. This builds on work of previous authors \citep[e.g.][]{Yan+Lazarian+Draine_2004} by taking a numerical (rather than approximate analytic) approach and probing regimes not considered previously.}
    \item {Lorentz forces, surprisingly, seem to offer grains relative protection from shattering via a greater degree of coupling to the underlying fluid.}
    \item We find that large grains should have relatively high erosion rates due to collisions with smaller grains, giving a typical lifetime for a micron-sized grain between 2.5 and 250 Myr, which is seemingly inconsistent with studies of similarly sized presolar grains found trapped in meteorites \citep{Takigawa_2018, Heck_2020}. 
\end{itemize}


\section*{Data availability}
The data presented in this paper was generated on the Princeton computing cluster `Stellar' and will be made freely available upon request to the corresponding author.
\section*{Acknowledgements}
We thank Matt Kunz and Eliot Quataert for their role in the initial trajectory of this project. We also thank Christopher Bambic and Ben Israeli for engaging discussions on particle kinetics, and Ulrich Steinwandel for his thoughts on dusty fluid dynamics. The lead author would also like to thank Jonathan Squire and Phil Hopkins for their role in sparking his interest in dusty fluids. We also thank the anonymous reviewer for his/her helpful comments that significantly improved the quality of the paper.






\bibliographystyle{mnras}
\bibliography{ms_reduced}

\begin{thebibliography}{}
\makeatletter
\relax
\def\mn@urlcharsother{\let\do\@makeother \do\$\do\&\do\#\do\^\do\_\do\%\do\~}
\def\mn@doi{\begingroup\mn@urlcharsother \@ifnextchar [ {\mn@doi@}
  {\mn@doi@[]}}
\def\mn@doi@[#1]#2{\def\@tempa{#1}\ifx\@tempa\@empty \href
  {http://dx.doi.org/#2} {doi:#2}\else \href {http://dx.doi.org/#2} {#1}\fi
  \endgroup}
\def\mn@eprint#1#2{\mn@eprint@#1:#2::\@nil}
\def\mn@eprint@arXiv#1{\href {http://arxiv.org/abs/#1} {{\tt arXiv:#1}}}
\def\mn@eprint@dblp#1{\href {http://dblp.uni-trier.de/rec/bibtex/#1.xml}
  {dblp:#1}}
\def\mn@eprint@#1:#2:#3:#4\@nil{\def\@tempa {#1}\def\@tempb {#2}\def\@tempc
  {#3}\ifx \@tempc \@empty \let \@tempc \@tempb \let \@tempb \@tempa \fi \ifx
  \@tempb \@empty \def\@tempb {arXiv}\fi \@ifundefined
  {mn@eprint@\@tempb}{\@tempb:\@tempc}{\expandafter \expandafter \csname
  mn@eprint@\@tempb\endcsname \expandafter{\@tempc}}}

\bibitem[\protect\citeauthoryear{Anderl, Guillet, Des~For{\^e}ts  \&
  Flower}{Anderl et~al.}{2013}]{anderl2013shocks}
Anderl S.,  Guillet V.,  Des~For{\^e}ts G.~P.,   Flower D.,  2013, Astronomy \&
  Astrophysics, 556, A69

\bibitem[\protect\citeauthoryear{Anders \& Zinner}{Anders \&
  Zinner}{1993}]{Anders+Zinner_1993}
Anders E.,  Zinner E.,  1993, Meteoritics, 28, 490

\bibitem[\protect\citeauthoryear{Aoyama, Hirashita  \& Nagamine}{Aoyama
  et~al.}{2020}]{Aoyama+Hirashita_2020}
Aoyama S.,  Hirashita H.,   Nagamine K.,  2020, Monthly Notices of the Royal
  Astronomical Society, 491, 3844

\bibitem[\protect\citeauthoryear{Bai \& Stone}{Bai \&
  Stone}{2010}]{bai2010particle}
Bai X.-N.,  Stone J.~M.,  2010, The Astrophysical Journal Supplement Series,
  190, 297

\bibitem[\protect\citeauthoryear{Bai, Caprioli, Sironi  \& Spitkovsky}{Bai
  et~al.}{2015}]{bai2015magnetohydrodynamic}
Bai X.-N.,  Caprioli D.,  Sironi L.,   Spitkovsky A.,  2015, The Astrophysical
  Journal, 809, 55

\bibitem[\protect\citeauthoryear{Bai, Ostriker, Plotnikov  \& Stone}{Bai
  et~al.}{2019}]{bai2019magnetohydrodynamic}
Bai X.-N.,  Ostriker E.~C.,  Plotnikov I.,   Stone J.~M.,  2019, The
  Astrophysical Journal, 876, 60

\bibitem[\protect\citeauthoryear{{Baines}, {Williams}  \& {Asebiomo}}{{Baines}
  et~al.}{1965}]{Baines+Williams+Asebiomo_1965}
{Baines} M.~J.,  {Williams} I.~P.,   {Asebiomo} A.~S.,  1965, \mnras, 130, 63

\bibitem[\protect\citeauthoryear{Beitia-Antero \& G{\'o}mez~de
  Castro}{Beitia-Antero \& G{\'o}mez~de Castro}{2021}]{Beitia+Gomez_2021}
Beitia-Antero L.,  G{\'o}mez~de Castro A.~I.,  2021, The Astrophysical Journal,
  909, 206

\bibitem[\protect\citeauthoryear{Beitia-Antero, G{\'o}mez~de Castro  \&
  Vallejo}{Beitia-Antero et~al.}{2021}]{Beitia+Gomez+Vallejo_2021}
Beitia-Antero L.,  G{\'o}mez~de Castro A.~I.,   Vallejo J.~C.,  2021, The
  Astrophysical Journal, 908, 112

\bibitem[\protect\citeauthoryear{Bernstein, Freedman  \& Madore}{Bernstein
  et~al.}{2002}]{bernstein2002first}
Bernstein R.~A.,  Freedman W.~L.,   Madore B.~F.,  2002, The Astrophysical
  Journal, 571, 56

\bibitem[\protect\citeauthoryear{Berruyer}{Berruyer}{1991}]{berruyer1991dust}
Berruyer N.,  1991, Astronomy and Astrophysics, 249, 181

\bibitem[\protect\citeauthoryear{Boris}{Boris}{1970}]{boris1970relativistic}
Boris J.,  1970, in 4th conference on numerical simulation of plasma. p.~3

\bibitem[\protect\citeauthoryear{Carballido, Stone  \& Turner}{Carballido
  et~al.}{2008}]{carballido2008kinematics}
Carballido A.,  Stone J.~M.,   Turner N.~J.,  2008, Monthly Notices of the
  Royal Astronomical Society, 386, 145

\bibitem[\protect\citeauthoryear{Carrera \& Simon}{Carrera \&
  Simon}{2022}]{Carrera+Simon_2022}
Carrera D.,  Simon J.~B.,  2022, The Astrophysical Journal Letters, 933, L10

\bibitem[\protect\citeauthoryear{Chang, Schiano  \& Wolfe}{Chang
  et~al.}{1987}]{chang1987effect}
Chang C.,  Schiano A.,   Wolfe A.,  1987, The Astrophysical Journal, 322, 180

\bibitem[\protect\citeauthoryear{{Chiang} \& {Youdin}}{{Chiang} \&
  {Youdin}}{2010}]{Chiang+Youdin_2010}
{Chiang} E.,  {Youdin} A.~N.,  2010, \mn@doi [Annual Review of Earth and
  Planetary Sciences] {10.1146/annurev-earth-040809-152513}, \href
  {https://ui.adsabs.harvard.edu/abs/2010AREPS..38..493C} {38, 493}

\bibitem[\protect\citeauthoryear{Chokshi, Tielens  \& Hollenbach}{Chokshi
  et~al.}{1993}]{Chokshi+Tielens+Hollenbach_1993}
Chokshi A.,  Tielens A.,   Hollenbach D.,  1993, The Astrophysical Journal,
  407, 806

\bibitem[\protect\citeauthoryear{Di~Mascia, Gallerani, Ferrara, Pallottini,
  Maiolino, Carniani  \& D’Odorico}{Di~Mascia et~al.}{2021}]{Di_2021}
Di~Mascia F.,  Gallerani S.,  Ferrara A.,  Pallottini A.,  Maiolino R.,
  Carniani S.,   D’Odorico V.,  2021, Monthly Notices of the Royal
  Astronomical Society, 506, 3946

\bibitem[\protect\citeauthoryear{Dom{\'\i}nguez-Fern{\'a}ndez, Br{\"u}ggen,
  Vazza, Banda-Barrag{\'a}n, Rajpurohit, Mignone, Mukherjee  \&
  Vaidya}{Dom{\'\i}nguez-Fern{\'a}ndez et~al.}{2021}]{dominguez2021morphology}
Dom{\'\i}nguez-Fern{\'a}ndez P.,  Br{\"u}ggen M.,  Vazza F.,
  Banda-Barrag{\'a}n W.,  Rajpurohit K.,  Mignone A.,  Mukherjee D.,   Vaidya
  B.,  2021, Monthly Notices of the Royal Astronomical Society, 500, 795

\bibitem[\protect\citeauthoryear{{Draine}}{{Draine}}{2003}]{Draine_2003a}
{Draine} B.~T.,  2003, \mn@doi [\araa]
  {10.1146/annurev.astro.41.011802.094840}, \href
  {http://adsabs.harvard.edu/abs/2003ARA%26A..41..241D} {41, 241}

\bibitem[\protect\citeauthoryear{{Draine}}{{Draine}}{2011}]{Draine_2011a}
{Draine} B.~T.,  2011, {Physics of the Interstellar and Intergalactic Medium}.
Princeton, NJ: Princeton Univ.\ Press

\bibitem[\protect\citeauthoryear{{Draine} \& {Salpeter}}{{Draine} \&
  {Salpeter}}{1979}]{Draine+Salpeter_1979a}
{Draine} B.~T.,  {Salpeter} E.~E.,  1979, \mn@doi [\apj] {10.1086/157165},
  \href {http://adsabs.harvard.edu/abs/1979ApJ...231...77D} {231, 77}

\bibitem[\protect\citeauthoryear{Epstein}{Epstein}{1924}]{epstein1924resistance}
Epstein P.~S.,  1924, Physical Review, 23, 710

\bibitem[\protect\citeauthoryear{{Fermi}}{{Fermi}}{1949}]{Fermi_1949}
{Fermi} E.,  1949, \mn@doi [\pr] {10.1103/PhysRev.75.1169}, 75, 1169

\bibitem[\protect\citeauthoryear{Franco, Ferrini, Ferrara  \& Barsella}{Franco
  et~al.}{1991}]{franco1991photolevitation}
Franco J.,  Ferrini F.,  Ferrara A.,   Barsella B.,  1991, The Astrophysical
  Journal, 366, 443

\bibitem[\protect\citeauthoryear{Fromang, Hennebelle  \& Teyssier}{Fromang
  et~al.}{2006}]{fromang2006high}
Fromang S.,  Hennebelle P.,   Teyssier R.,  2006, Astronomy \& Astrophysics,
  457, 371

\bibitem[\protect\citeauthoryear{Gail \& Sedlmayr}{Gail \&
  Sedlmayr}{1999}]{gail1999mineral}
Gail H.-P.,  Sedlmayr E.,  1999, Astronomy and Astrophysics, 347, 594

\bibitem[\protect\citeauthoryear{Gehrz}{Gehrz}{1989}]{gehrz1989sources}
Gehrz R.~D.,  1989, in Symposium-International astronomical union. pp 445--453

\bibitem[\protect\citeauthoryear{Guillet, Des~For{\^e}ts  \& Jones}{Guillet
  et~al.}{2007}]{guillet2007shocks}
Guillet V.,  Des~For{\^e}ts G.~P.,   Jones A.,  2007, Astronomy \&
  Astrophysics, 476, 263

\bibitem[\protect\citeauthoryear{Guillet, Jones  \& Des~For{\^e}ts}{Guillet
  et~al.}{2009}]{guillet2009shocks}
Guillet V.,  Jones A.,   Des~For{\^e}ts G.~P.,  2009, Astronomy \&
  Astrophysics, 497, 145

\bibitem[\protect\citeauthoryear{Guillet, Des~For{\^e}ts  \& Jones}{Guillet
  et~al.}{2011}]{guillet2011shocks}
Guillet V.,  Des~For{\^e}ts G.~P.,   Jones A.,  2011, Astronomy \&
  Astrophysics, 527, A123

\bibitem[\protect\citeauthoryear{{Haugen}, {Brandenburg}, {Sandin}  \&
  {Mattsson}}{{Haugen} et~al.}{2022}]{Haugen+Brandenburg+Mattson_2022}
{Haugen} N. E.~L.,  {Brandenburg} A.,  {Sandin} C.,   {Mattsson} L.,  2022,
  \mn@doi [Journal of Fluid Mechanics] {10.1017/jfm.2021.1143}, \href
  {https://ui.adsabs.harvard.edu/abs/2022JFM...934A..37H} {934, A37}

\bibitem[\protect\citeauthoryear{Heck et~al.,}{Heck et~al.}{2020}]{Heck_2020}
Heck P.~R.,  et~al., 2020, Proceedings of the National Academy of Sciences,
  117, 1884

\bibitem[\protect\citeauthoryear{Hirashita \& Ferrara}{Hirashita \&
  Ferrara}{2002}]{hirashita2002effects}
Hirashita H.,  Ferrara A.,  2002, Monthly Notices of the Royal Astronomical
  Society, 337, 921

\bibitem[\protect\citeauthoryear{{Hirashita} \& {Yan}}{{Hirashita} \&
  {Yan}}{2009}]{Hirashita+Yan_2009}
{Hirashita} H.,  {Yan} H.,  2009, \mn@doi [\mnras]
  {10.1111/j.1365-2966.2009.14405.x}, 394, 1061

\bibitem[\protect\citeauthoryear{Hirashita, Hunt  \& Ferrara}{Hirashita
  et~al.}{2002}]{Hirashita+Hunt+Ferrara_2002}
Hirashita H.,  Hunt L.~K.,   Ferrara A.,  2002, Monthly Notices of the Royal
  Astronomical Society, 330, L19

\bibitem[\protect\citeauthoryear{{Hoang}, {Lazarian}  \&
  {Schlickeiser}}{{Hoang} et~al.}{2012}]{Hoang+Lazarian+Schlickeiser_2012}
{Hoang} T.,  {Lazarian} A.,   {Schlickeiser} R.,  2012, \mn@doi [\apj]
  {10.1088/0004-637X/747/1/54}, \href
  {http://adsabs.harvard.edu/abs/2012ApJ...747...54H} {747, 54}

\bibitem[\protect\citeauthoryear{H{\"o}fner \& Olofsson}{H{\"o}fner \&
  Olofsson}{2018}]{hofner2018mass}
H{\"o}fner S.,  Olofsson H.,  2018, The Astronomy and Astrophysics Review, 26,
  1

\bibitem[\protect\citeauthoryear{Hopkins}{Hopkins}{2015}]{Hopkins_2015}
Hopkins P.~F.,  2015, Monthly Notices of the Royal Astronomical Society, 450,
  53

\bibitem[\protect\citeauthoryear{Hopkins \& Lee}{Hopkins \&
  Lee}{2016}]{hopkins2016fundamentally}
Hopkins P.~F.,  Lee H.,  2016, Monthly Notices of the Royal Astronomical
  Society, 456, 4174

\bibitem[\protect\citeauthoryear{Hopkins \& Squire}{Hopkins \&
  Squire}{2018a}]{hopkins2018ubiquitous}
Hopkins P.~F.,  Squire J.,  2018a, Monthly Notices of the Royal Astronomical
  Society, 479, 4681

\bibitem[\protect\citeauthoryear{Hopkins \& Squire}{Hopkins \&
  Squire}{2018b}]{hopkins2018resonant}
Hopkins P.~F.,  Squire J.,  2018b, Monthly Notices of the Royal Astronomical
  Society, 480, 2813

\bibitem[\protect\citeauthoryear{Johansen, Youdin  \& Mac~Low}{Johansen
  et~al.}{2009}]{johansen2009particle}
Johansen A.,  Youdin A.,   Mac~Low M.-M.,  2009, The Astrophysical Journal,
  704, L75

\bibitem[\protect\citeauthoryear{Johansen et~al.,}{Johansen
  et~al.}{2014}]{Johansen_2014}
Johansen A.,  et~al., 2014, arXiv preprint arXiv:1402.1344, pp 547--548

\bibitem[\protect\citeauthoryear{{Jones}, {Tielens}  \& {Hollenbach}}{{Jones}
  et~al.}{1996}]{Jones+Tielens+Hollenbach_1996}
{Jones} A.~P.,  {Tielens} A.~G.~G.~M.,   {Hollenbach} D.~J.,  1996, \mn@doi
  [\apj] {10.1086/177823}, 469, 740

\bibitem[\protect\citeauthoryear{Kwok}{Kwok}{1975}]{kwok1975radiation}
Kwok S.,  1975, The Astrophysical Journal, 198, 583

\bibitem[\protect\citeauthoryear{Laibe \& Price}{Laibe \&
  Price}{2012a}]{laibe2012dusty}
Laibe G.,  Price D.~J.,  2012a, Monthly Notices of the Royal Astronomical
  Society, 420, 2345

\bibitem[\protect\citeauthoryear{Laibe \& Price}{Laibe \&
  Price}{2012b}]{laibe2012dustyII}
Laibe G.,  Price D.~J.,  2012b, Monthly Notices of the Royal Astronomical
  Society, 420, 2365

\bibitem[\protect\citeauthoryear{Laibe \& Price}{Laibe \&
  Price}{2014}]{laibe2014dusty}
Laibe G.,  Price D.~J.,  2014, Monthly Notices of the Royal Astronomical
  Society, 440, 2136

\bibitem[\protect\citeauthoryear{{Lamers} \& {Cassinelli}}{{Lamers} \&
  {Cassinelli}}{1999}]{Lamers_Cassinelli_1999}
{Lamers} H. J. G. L.~M.,  {Cassinelli} J.~P.,  1999, Introduction to Stellar
  Winds.
Cambridge University Press, Cambridge, U. K.

\bibitem[\protect\citeauthoryear{{Larson}}{{Larson}}{1981}]{Larson_1981}
{Larson} R.~B.,  1981, \mnras, 194, 809

\bibitem[\protect\citeauthoryear{{Lazarian} \& {Yan}}{{Lazarian} \&
  {Yan}}{2002}]{Lazarian+Yan_2002}
{Lazarian} A.,  {Yan} H.,  2002, \mn@doi [\apjl] {10.1086/339675}, \href
  {http://adsabs.harvard.edu/abs/2002ApJ...566L.105L} {566, L105}

\bibitem[\protect\citeauthoryear{Lebreuilly, Commer{\c{c}}on  \&
  Laibe}{Lebreuilly et~al.}{2019}]{lebreuilly2019small}
Lebreuilly U.,  Commer{\c{c}}on B.,   Laibe G.,  2019, Astronomy \&
  Astrophysics, 626, A96

\bibitem[\protect\citeauthoryear{Lee, Hopkins  \& Squire}{Lee
  et~al.}{2017}]{lee2017dynamics}
Lee H.,  Hopkins P.~F.,   Squire J.,  2017, Monthly Notices of the Royal
  Astronomical Society, 469, 3532

\bibitem[\protect\citeauthoryear{{Mathis}, {Rumpl}  \& {Nordsieck}}{{Mathis}
  et~al.}{1977}]{Mathis+Rumpl+Nordsieck_1977}
{Mathis} J.~S.,  {Rumpl} W.,   {Nordsieck} K.~H.,  1977, \apj, 217, 425

\bibitem[\protect\citeauthoryear{{Mattsson} \& {Hedvall}}{{Mattsson} \&
  {Hedvall}}{2022}]{Mattson+Hedvall_2022}
{Mattsson} L.,  {Hedvall} R.,  2022, \mn@doi [\mnras] {10.1093/mnras/stab3216},
  \href {https://ui.adsabs.harvard.edu/abs/2022MNRAS.509.3660M} {509, 3660}

\bibitem[\protect\citeauthoryear{Melands{\o} \& Shukla}{Melands{\o} \&
  Shukla}{1995}]{melandso1995theory}
Melands{\o} F.,  Shukla P.,  1995, Planetary and Space Science, 43, 635

\bibitem[\protect\citeauthoryear{Mignone, Bodo, Vaidya  \& Mattia}{Mignone
  et~al.}{2018}]{mignone2018particle}
Mignone A.,  Bodo G.,  Vaidya B.,   Mattia G.,  2018, The Astrophysical
  Journal, 859, 13

\bibitem[\protect\citeauthoryear{Moseley, Squire  \& Hopkins}{Moseley
  et~al.}{2019}]{moseley2019non}
Moseley E.~R.,  Squire J.,   Hopkins P.~F.,  2019, Monthly Notices of the Royal
  Astronomical Society, 489, 325

\bibitem[\protect\citeauthoryear{Moseley, Draine, Tomida  \& Stone}{Moseley
  et~al.}{2021}]{Moseley+Draine+Tomida_2021}
Moseley E.~R.,  Draine B.,  Tomida K.,   Stone J.~M.,  2021, Monthly Notices of
  the Royal Astronomical Society, 500, 3290

\bibitem[\protect\citeauthoryear{{Moseley}, {Teyssier}  \& {Draine}}{{Moseley}
  et~al.}{2022}]{MoseleyInPrep}
{Moseley} E.~R.,  {Teyssier} R.,   {Draine} B.~T.,  2022, in preparation

\bibitem[\protect\citeauthoryear{Orszag \& Tang}{Orszag \&
  Tang}{1979}]{orszag1979small}
Orszag S.~A.,  Tang C.-M.,  1979, Journal of Fluid Mechanics, 90, 129

\bibitem[\protect\citeauthoryear{Padoan, Jimenez, Juvela  \& Nordlund}{Padoan
  et~al.}{2004}]{padoan2004average}
Padoan P.,  Jimenez R.,  Juvela M.,   Nordlund {\AA}.,  2004, The Astrophysical
  Journal, 604, L49

\bibitem[\protect\citeauthoryear{Pan, Padoan, Scalo, Kritsuk  \& Norman}{Pan
  et~al.}{2011}]{pan2011turbulent}
Pan L.,  Padoan P.,  Scalo J.,  Kritsuk A.~G.,   Norman M.~L.,  2011, The
  Astrophysical Journal, 740, 6

\bibitem[\protect\citeauthoryear{Plane}{Plane}{2013}]{plane2013nucleation}
Plane J.~M.,  2013, Philosophical Transactions of the Royal Society A:
  Mathematical, Physical and Engineering Sciences, 371, 20120335

\bibitem[\protect\citeauthoryear{Sandford, Whitaker  \& Klein}{Sandford
  et~al.}{1984}]{sandford1984radiatively}
Sandford M.,  Whitaker R.,   Klein R.,  1984, The Astrophysical Journal, 282,
  178

\bibitem[\protect\citeauthoryear{Schilke, Walmsley, Pineau~des Forets  \&
  Flower}{Schilke et~al.}{1997}]{schilke1997sio}
Schilke P.,  Walmsley C.,  Pineau~des Forets G.,   Flower D.,  1997, Astronomy
  and Astrophysics, 321, 293

\bibitem[\protect\citeauthoryear{Seligman, Hopkins  \& Squire}{Seligman
  et~al.}{2019}]{seligman2019non}
Seligman D.,  Hopkins P.~F.,   Squire J.,  2019, Monthly Notices of the Royal
  Astronomical Society, 485, 3991

\bibitem[\protect\citeauthoryear{Soifer et~al.,}{Soifer
  et~al.}{1984}]{soifer1984infrared}
Soifer B.,  et~al., 1984, The Astrophysical Journal, 278, L71

\bibitem[\protect\citeauthoryear{{Solomon}, {Rivolo}, {Barrett}  \&
  {Yahil}}{{Solomon} et~al.}{1987}]{Solomon+Rivolo+Barrett+Yahil_1987}
{Solomon} P.~M.,  {Rivolo} A.~R.,  {Barrett} J.,   {Yahil} A.,  1987, \mn@doi
  [\apj] {10.1086/165493}, 319, 730

\bibitem[\protect\citeauthoryear{{Spitzer}}{{Spitzer}}{1978}]{Spitzer_1978}
{Spitzer} L.,  1978, Physical Processes in the Interstellar Medium.
Wiley, New York

\bibitem[\protect\citeauthoryear{Squire \& Hopkins}{Squire \&
  Hopkins}{2018}]{squire2018letter}
Squire J.,  Hopkins P.~F.,  2018, The Astrophysical Journal Letters, 856, L15

\bibitem[\protect\citeauthoryear{Steinwandel, Kaurov, Hopkins  \&
  Squire}{Steinwandel et~al.}{2021}]{steinwandel2021optical}
Steinwandel U.~P.,  Kaurov A.~A.,  Hopkins P.~F.,   Squire J.,  2021, arXiv
  preprint arXiv:2111.09335

\bibitem[\protect\citeauthoryear{Takigawa, Stroud, Nittler, Alexander  \&
  Miyake}{Takigawa et~al.}{2018}]{Takigawa_2018}
Takigawa A.,  Stroud R.~M.,  Nittler L.~R.,  Alexander C.~M.,   Miyake A.,
  2018, The Astrophysical Journal Letters, 862, L13

\bibitem[\protect\citeauthoryear{Teyssier}{Teyssier}{2002}]{teyssier2002cosmological}
Teyssier R.,  2002, Astronomy \& Astrophysics, 385, 337

\bibitem[\protect\citeauthoryear{Teyssier, Fromang  \& Dormy}{Teyssier
  et~al.}{2006}]{teyssier2006kinematic}
Teyssier R.,  Fromang S.,   Dormy E.,  2006, Journal of Computational Physics,
  218, 44

\bibitem[\protect\citeauthoryear{{Thompson}, {Quataert}  \&
  {Murray}}{{Thompson} et~al.}{2005}]{Thompson+Quataert+Murray_2005}
{Thompson} T.~A.,  {Quataert} E.,   {Murray} N.,  2005, \mn@doi [\apj]
  {10.1086/431923}, 630, 167

\bibitem[\protect\citeauthoryear{Todini \& Ferrara}{Todini \&
  Ferrara}{2001}]{Todini+Ferrara_2001}
Todini P.,  Ferrara A.,  2001, Monthly Notices of the Royal Astronomical
  Society, 325, 726

\bibitem[\protect\citeauthoryear{Tricco, Price  \& Laibe}{Tricco
  et~al.}{2017}]{tricco2017dust}
Tricco T.~S.,  Price D.~J.,   Laibe G.,  2017, Monthly Notices of the Royal
  Astronomical Society: Letters, 471, L52

\bibitem[\protect\citeauthoryear{{Trumpler}}{{Trumpler}}{1930}]{Trumpler_1930}
{Trumpler} R.~J.,  1930, \pasp, 42, 214

\bibitem[\protect\citeauthoryear{Van~Leer}{Van~Leer}{1979}]{VanLeer_1979}
Van~Leer B.,  1979, Journal of computational Physics, 32, 101

\bibitem[\protect\citeauthoryear{{Weingartner} \& {Draine}}{{Weingartner} \&
  {Draine}}{1999}]{Weingartner+Draine_1999}
{Weingartner} J.~C.,  {Draine} B.~T.,  1999, \mn@doi [\apj] {10.1086/307197},
  \href {http://adsabs.harvard.edu/abs/1999ApJ...517..292W} {517, 292}

\bibitem[\protect\citeauthoryear{{Weingartner} \& {Draine}}{{Weingartner} \&
  {Draine}}{2001}]{Weingartner+Draine_2001a}
{Weingartner} J.~C.,  {Draine} B.~T.,  2001, \mn@doi [\apj] {10.1086/318651},
  \href {http://adsabs.harvard.edu/abs/2001ApJ...548..296W} {548, 296}

\bibitem[\protect\citeauthoryear{Yan}{Yan}{2009}]{Yan_2009}
Yan H.,  2009, Monthly Notices of the Royal Astronomical Society, 397, 1093

\bibitem[\protect\citeauthoryear{{Yan} \& {Lazarian}}{{Yan} \&
  {Lazarian}}{2003}]{Yan+Lazarian_2003}
{Yan} H.,  {Lazarian} A.,  2003, \mn@doi [\apjl] {10.1086/377487}, \href
  {http://adsabs.harvard.edu/abs/2003ApJ...592L..33Y} {592, L33}

\bibitem[\protect\citeauthoryear{{Yan}, {Lazarian}  \& {Draine}}{{Yan}
  et~al.}{2004}]{Yan+Lazarian+Draine_2004}
{Yan} H.,  {Lazarian} A.,   {Draine} B.~T.,  2004, \mn@doi [\apj]
  {10.1086/425111}, 616, 895

\bibitem[\protect\citeauthoryear{Yang \& Johansen}{Yang \&
  Johansen}{2016}]{yang2016integration}
Yang C.-C.,  Johansen A.,  2016, The Astrophysical Journal Supplement Series,
  224, 39

\bibitem[\protect\citeauthoryear{Yang, Johansen  \& Carrera}{Yang
  et~al.}{2017}]{yang2017concentrating}
Yang C.-C.,  Johansen A.,   Carrera D.,  2017, Astronomy \& Astrophysics, 606,
  A80

\bibitem[\protect\citeauthoryear{Youdin \& Goodman}{Youdin \&
  Goodman}{2005}]{youdin2005streaming}
Youdin A.~N.,  Goodman J.,  2005, The Astrophysical Journal, 620, 459

\bibitem[\protect\citeauthoryear{Youdin \& Johansen}{Youdin \&
  Johansen}{2007}]{youdin2007protoplanetary}
Youdin A.,  Johansen A.,  2007, The Astrophysical Journal, 662, 613

\bibitem[\protect\citeauthoryear{Zhuravlev}{Zhuravlev}{2021}]{zhuravlev2021dynamic}
Zhuravlev V.,  2021, Monthly Notices of the Royal Astronomical Society, 500,
  2209

\makeatother
\end{thebibliography}



\appendix
\end{document}